\documentclass[fleqn,usenatbib]{mnras}

\usepackage[T1]{fontenc}
\usepackage{ae,aecompl}

\DeclareRobustCommand{\VAN}[3]{#2}
\let\VANthebibliography\thebibliography
\def\thebibliography{\DeclareRobustCommand{\VAN}[3]{##3}\VANthebibliography}

\usepackage{graphicx}	
\usepackage{amsmath}	
\usepackage{amssymb}	
\usepackage{times}

\usepackage{newtxtext,newtxmath}

\newcommand{\gramses}{\textsc{gramses}}
\newcommand{\dtfe}{\textsc{dtfe}}
\newcommand{\camb}{\textsc{camb}}

\def\Mpch{~h^{-1} {\rm Mpc}}
\def\Gpch{~h^{-1} {\rm Gpc}}
\def\invMpch{~h{\rm Mpc}^{-1}}
\def\kpch{~h^{-1} {\rm kpc}}

\newcommand{\LCDM}{$\Lambda$CDM}
\def\B{{\bf B}}

\title[Gravitomagnetic effects in cross-correlations]{Looking for a twist: probing the cosmological gravitomagnetic effect via weak lensing-kSZ cross correlations}

\author[C. Barrera-Hinojosa et al.]{
Cristian Barrera-Hinojosa$^{1}$\thanks{E-mail: cristian.g.barrera@durham.ac.uk (CB-H)},
Baojiu Li$^{1}$
and Yan-Chuan Cai$^{2}$
\\
$^{1}$Institute for Computational Cosmology, Department of Physics, Durham University,
Durham DH1 3LE, UK\\
$^{2}$Institute for Astronomy, University of Edinburgh, Royal Observatory, Blackford Hill, Edinburgh EH9 3HJ, UK
}

\date{Accepted XXX. Received YYY; in original form ZZZ}

\pubyear{2021}

\begin{document}
\label{firstpage}
\pagerange{\pageref{firstpage}--\pageref{lastpage}}
\maketitle

\begin{abstract}
General relativity predicts that the rotational momentum flux of matter twists the spacetime via a vector gravitomagnetic (frame-dragging) field, which remains undetected in cosmology. This vector field induces an additional gravitational lensing effect; at the same time, the momentum field sources the kinetic Sunyaev-Zel'dovich (kSZ) effect. The common origin of these two effects allows us to probe the gravitomagnetic signal via their cross-correlations. In this paper, we explore the possibility of detecting the gravitomagnetic field in \LCDM{} by cross-correlating the weak-lensing convergence field with the CMB temperature map, which is imprinted with the kSZ signal. This approach allows us to extract the gravitomagnetic effect because the cross correlation between the standard Newtonian contribution to the weak-lensing convergence field, $\kappa_\Phi$, and the kSZ effect is expected to vanish. We study the cross correlations with a suite of large-volume Newtonian $N$-body simulations and a small-volume, high-resolution, general-relativistic counterpart. 
We show that insufficient simulation resolution can introduce significant spurious correlations between $\kappa_\Phi$ and kSZ. From the high-resolution simulation, we find that the cumulative signal-to-noise ratio (SNR) of the kSZ-gravitomagnetic convergence field can reach almost 15 (30) at $\ell\simeq5000$ ($10^4$) for the lensing source redshift $z_s=0.83$, if only cosmic variance is considered. We make forecast for next-generation lensing surveys such as \textsc{euclid} and \textsc{lsst}, and CMB experiments such as Simons Observatory and \textsc{cmb}-\textsc{s4}, and find that, for $z_s=1.4$, the cumulative SNR can exceed 5 (9) at $\ell\simeq5000$ ($10^4$), indicating that the cosmological gravitomagnetic effect can be detected, if several foreground contaminations can be removed.
\end{abstract}




\section{Introduction}

In General Relativity (GR), the propagation of light can be distorted not only by the Newtonian (scalar) potential, but also by the vector (spin-1) and tensor (spin-2) degrees of freedom of the gravitational field.
The leading-order post-Newtonian correction to Newtonian gravity corresponds to the gravitomagnetic (frame-dragging) potential --- a vector-type perturbation of the gravitational field that describes the twisting of the spacetime due to rotational matter flows. Its effects within the Solar System have been detected in the last two decades by Gravity Probe B~\citep{Everitt:2011hp}, but its faint cosmological signal is swamped by the Newtonian signal. With the advent of various upcoming large-scale structure surveys such as \textsc{euclid} \citep{euclid}, \textsc{lsst} \citep{lsst} and \textsc{ska} \citep{ska}, a renewed interest to understand in detail the impact of the vector potential on observables has emerged in recent years \citep[e.g.,][]{Schaefer:2005up,Andrianomena:2014sya,sagaWeakLensingInduced2015,Thomas:2015dfa,Cuesta-Lazaro:2018uyv,Tang:2020com}. There has also been growing interest in the gravitomagnetic effects on smaller astronomical systems. For instance, it has been argued that the observed flat rotation curves of galaxies potentially admits an alternative explanation --- in the absence of dark matter -- by a GR velocity profile sourced by frame-dragging \citep{Crosta:2018var}, although a realistic model for this is still required. Furthermore, a mission specially designed to measure the gravitomagnetic field of the Milky Way, and of its dark matter halo, has been recently proposed in \citet{tartagliaDetectingGravitomagneticField2021}.

Although vector modes are not introduced by the standard inflationary model~\citep[e.g.,][]{Bassett:2005xm}, the late-time gravitomagnetic potential of the $\Lambda$-cold-dark-matter (\LCDM{}) cosmology is generated dynamically: before shell crossing, this is sourced by the coupling of scalar perturbations of the matter fluid --- the overdensity and velocity divergence fields --- and hence this is typically referred to as the scalar-induced cosmological vector mode~\citep{Matarrese:1994wa,Lu:2007cj,Lu:2008ju}. 
More generally, the gravitomagnetic field is sourced and sustained over time by the rotational (divergence-free) component of the momentum flux of matter, and the latter also receives contributions from the vorticity field generated due to, e.g., shell crossing of CDM. As shown in \citet{Lu:2007cj,Lu:2008ju}, second-order perturbation theory predicts that, on scales above the matter-radiation equality scale (i.e., the horizon scale at the time of matter-radiation equality), the power spectrum of the gravitomagnetic field is strongly suppressed with respect to the Newtonian potential, but on sub-equality scales the relative amplitude can reach about $1\%$, which is also supported by $N$-body simulations~\citep{Thomas:2014aga, Adamek:2015eda, Barrera-Hinojosa:2020gnx}.

Even though the effect of the gravitomagnetic force in cosmological structure formation is {small due to the} low velocities of non-relativistic matter \citep{Adamek:2015eda,Barrera-Hinojosa:2020gnx}, which is at most of order $\mathcal{O}(1\%)$ of the speed of light, it is not {\it a priori} obvious that the impact on observations is negligible, since this requires to quantify the effect on the propagation of photons. However, so far all investigations have found that the gravitomagnetic effects in light propagation are subdominant with respect to their Newtonian (scalar) counterparts. For instance, it has been shown that the corrections to the observed galaxy number counts induced by the vector modes are too small to be detected by the upcoming surveys~\citep{Durrer:2016jzq}. Similarly, the second-order gravitomagnetic corrections to the lensing convergence field have also been found to have an overall negligible impact in most cases \citep[][]{Schaefer:2005up,Thomas:2014aga,Cuesta-Lazaro:2018uyv}, although these can still dominate over other relativistic effects in surveys with {\sc ska}-like source distributions \citep{Andrianomena:2014sya}. 

In the context of lensing, B-modes represent a characteristic signal imprinted by vector perturbations that, in principle, might be used to disentangle these from the effects of scalar perturbations (although B-modes are also induced by tensor perturbations, i.e., gravitational waves, their contribution is subdominant). However, as shown by \citet{sagaWeakLensingInduced2015}, a detection of the B-modes is not within the reach of upcoming galaxy surveys, although it has been argued that the large volume covered by future 21cm observations could improve the signal-to-noise ratio. In the same spirit, \citet{Tang:2020com} has recently proposed an estimator to measure the dipole feature in the lensing convergence field that is induced by the stacked rotation of clusters, although they predict that this signal is unlikely to be detected by \textsc{lsst}.

As originally suggested by \citet{Schaefer:2005up}, a potentially promising and yet unexplored way to extract the gravitomagnetic effects from lensing observations is via the cross correlation with a second observable, in particular with the kinetic Sunyaev-Zel'dovich (kSZ) effect  \citep{kSZ:1980,Ostriker-Vishniac:1986}. The kSZ effect is a secondary CMB anisotropy induced by the scattering of CMB photons off fast-moving free electrons in the intergalactic medium.
This particular signal is chosen because, just like the gravitomagnetic field, it is sourced by the momentum field of matter. More precisely, on small angular scales --- where the kSZ effect dominates over the primary CMB --- only rotational modes of the momentum field of matter will survive during the line-of-sight integration and contribute to this effect \citep[e.g.,][]{Zhang:2003nr}. Hence, the cross correlation between the kSZ effect and the gravitomagnetic convergence field is roughly proportional to the auto-correlation of either of the two effects. Furthermore, the kSZ effect is uncorrelated with the Newtonian (scalar) weak-lensing signal at the two-point level due to the statistical isotropy of the velocity field \citep{Dore:2003ex}, making it an ideal probe to extract the gravitomagnetic (vector) contribution of the convergence field.  

In this paper we will explore, for the first time, the detectability of the cosmological gravitomagnetic field via cross correlation of the weak-lensing convergence field --- that contains both Newtonian and gravitomagnetic contributions --- and the kSZ effect. Because in practice it is not always easy to separate the kSZ effect from the primary CMB, we shall consider the cross correlation between the total lensing convergence and a total CMB temperature map, the latter including the kSZ effect integrated over lines of sight. We will also discuss the impact of other secondary CMB anisotropies on this cross correlation.

The outline of the remainder of this paper is as follows. In Section \ref{sec:theory} we discuss the key theoretical aspects of the gravitomagnetic contribution to the weak-lensing convergence field, its angular power spectrum and the convergence-kSZ cross angular power spectrum. In Section \ref{sec:simulations} we present the details and specifications of the $N$-body simulations used to model the observables. In Section \ref{sec:modelling_obs} we describe the methodology that we use to generate the sky maps for the above observables, while we devote Section \ref{subsect:noise} to study in detail the unphysical (i.e., beyond the effect of cosmic variance) non-zero cross-correlation of kSZ and the scalar part of the convergence field found from the maps. 
Then, in Section \ref{sec:results} we present the main results of this paper, in which we quantify the signal-to-noise ratio of the gravitomagnetic signal based on a high-resolution simulation. 
In Section \ref{sec:detectability} we discuss the detectability of this signal with current and upcoming weak lensing surveys, such as \textsc{euclid} \citep{euclid} and Vera C.~Rubin Observatory \citep[\textsc{lsst};][]{lsst}, and CMB experiments including the Simons Observatory \citep{SimonsObservatory} and CMB Stage IV \citep[\textsc{cmb}-\textsc{s4};][]{CMBS4}. Finally, in Section \ref{sec:conclusion} we present our conclusions. 

\section{Theory}
\label{sec:theory}

In this paper, we consider a perturbed Friedmann-Lema\^itre-Robertson-Walker (FLRW) metric in the weak-field regime. In the Poisson (or longitudinal) gauge including scalar and vector modes, this is given by \citep{Ma:1995ey,Matarrese:1997ay}
\begin{align}
{\rm d}s^2=-\left(1+2\frac{\Phi}{c^2}\right)c^2{\rm d}t^2 + a^2\left(1-2\frac{\Psi}{c^2}\right){\rm d}{\bf x}^2 + 2a^2\frac{{\bf B}}{c^3}\cdot {\rm d}{\bf x}c{\rm d}t\,.
\label{eq:metric-general}
\end{align}
Here, $t$ is cosmic time, ${\bf x}$ are comoving spatial Cartesian coordinates, $a$ is the scale factor, $c$ is the speed of light, $\Phi$ and $\Psi$ are the scalar degrees of freedom corresponding to the Bardeen potentials, and $\B\equiv (B^x,B^y,B^z)$ is the gauge-invariant vector gravitomagnetic (frame-dragging) potential \citep{Bardeen:1980kt}, which satisfies the divergence-free (transverse) condition $\nabla\cdot\B=0$, where $\nabla$ denotes the derivative with respect to the comoving coordinates. Throughout this work we will neglect the gravitational slip and set $\Phi=\Psi$, which is identified as the Newtonian gravitational potential. On the other hand, in the weak-field approximation the matter fields such as density, velocity and momentum are treated as non-perturbative fields. 

The metric Eq.~\eqref{eq:metric-general} can also be obtained in a post-Newtonian (or more precisely, a Post-Friedmann) expansion up to leading order in $1/c^3$~\citep{schneiderGravitationalLenses1992,Sereno:2002tv}, which is valid at all scales~\citep{Bruni:2013mua,Milillo:2015cva}. In this approach, the dynamics of CDM is not modified by the presence of $\B$ at this order in the expansion, but observables are still affected through its effect on the photon geodesics, which is one of the main approximations assumed throughout this paper.

The Newtonian potential satisfies the Poisson equation
\begin{align}
    \nabla^2\Phi=\frac{3H^2_0\Omega_m}{2a}\delta\,,
 \label{eq:Poisson-equation}
\end{align}
where $\delta$ is the gauge-invariant density contrast, $H_0$ the Hubble constant, and $\Omega_m$ the present-day matter density parameter. 
The gravitomagnetic potential satisfies an analogue elliptic-type equation, in which the source term is the rotational component of the momentum density field. This is given by \citep{Bruni:2013mua} 
\begin{align}
 \frac{1}{4}  \nabla\times \nabla^2\B=\frac{3H^2_0\Omega_m}{2a}\nabla\times{\left[(1+\delta)\frac{{\bf v}}{c}\right]}\,,
 \label{eq:momentum-constraint}
\end{align}
where ${\bf q}=(1+\delta) {\bf v}$ is the momentum field of matter, ${\bf v}={\rm d}{\bf x}/{\rm d}t$ being the peculiar velocity. In Eq.~\eqref{eq:momentum-constraint}, the curl operator has been applied on both sides to remove the scalar component of the momentum field, which does not contribute to $\B$.
Note that Eq.~\eqref{eq:momentum-constraint} here has different $a$-factors compared to Eq.~(3) of \citet{Bruni:2013mua} due to the different conventions on the definition of $\B$, which can have either an upper or lower index, and of the peculiar velocity. The advantage of Eq.~\eqref{eq:momentum-constraint} is that, up to a factor of $1/4$, it has the identical form as  Eq.~\eqref{eq:Poisson-equation} apart from the matter source term, thus putting the two potentials on equal footing. Furthermore, this also offers a clear and compact way to write down the total lensing convergence field in the presence of gravitomagnetic effects, as discussed in the next subsection.

\subsection{The gravitomagnetic contributions to lensing convergence}

In the post-Newtonian regime, the total deflection angle of photons caused by a slowly moving perfect fluid can be obtained by replacing the standard lensing potential by an effective lensing potential given as \citep[e.g.,][]{schneiderGravitationalLenses1992,Sereno:2003tk,Schaefer:2005up}
\begin{equation}
\Phi \to \Phi+\frac{1}{2c}\B\cdot\hat{\bf n}\,,
\end{equation}
where $\hat{\bf n}$ is the unit vector of the line-of-sight (LOS) direction. Therefore, the total lensing convergence field can be written as
\begin{align}
\kappa_{\rm GR}(\hat{\bf n}) = \kappa_\Phi + \kappa_\B\,,
\label{eq:kappa_GR}
\end{align}
where the standard (Newtonian) and gravitomagnetic contributions to the convergence field are respectively given by
\begin{align}
    \kappa_\Phi(\hat{\bf n})&=\int{\rm d}\chi K_{\kappa_{\Phi}}(\chi)\delta(\chi\hat{\bf n},z)\,,\label{eq:kappa_Phi}\\
    \kappa_\B(\hat{\bf n})&=\int{\rm d}\chi K_{\kappa_\B}(\chi)[{\bf q}_\perp\cdot\hat{\bf n}](\chi\hat{\bf n},z)\,. \label{eq:kappa_B}
\end{align}
Here, $\chi$ is the comoving distance, ${\bf q}_\perp$ is the rotational (divergence-free) component of the momentum field, and $K_{\kappa_\Phi}$ is the standard weak-lensing kernel
\begin{equation}
    K_{\kappa_\Phi}(\chi) = \frac{3}{2}\frac{H_0^2\Omega_m}{ac^2}\int_0^{\chi}{\rm d}\chi'\frac{\chi(\chi'-\chi)}{\chi'}\frac{{\rm d}\chi'}{{\rm d}z}p\left(z(\chi')\right),
\label{eq:Kernel-lensing}
\end{equation}
while the gravitomagnetic lensing kernel satisfies \citep{Schaefer:2005up}
\begin{align}
K_{\kappa_\B}=\frac{2}{c}K_{\kappa_\Phi}\,.
\label{eq:Kernel-kappa-B}
\end{align}
In these, $p_z(z)$ is the normalised source redshift distribution, $\int {\rm d}z p_z(z)=1$, and the LOS integration is carried out up to the farthest source.
For definiteness, in this work we use a single source galaxy redshift $z_s$ (corresponding to a comoving distance $\chi_s$) with the source distribution given by
\begin{align}
    p_z(\chi(z))= \delta^{\rm D}(\chi - \chi_s)\,,
\end{align}
in which $\delta^{\rm D}$ is the Dirac $\delta$ function. In reality, $p_z(\chi)$ is a continuous distribution that depends on the specific galaxy survey used. 
Notice that in this post-Friedmann approximation, the gravitomagnetic convergence field, Eq. \eqref{eq:kappa_B}, is written in terms of the rotational modes of the momentum field using \eqref{eq:momentum-constraint}, just as it is customary to express $\kappa_\Phi$ in terms of the density field via \eqref{eq:Poisson-equation}.

Under the Limber approximation, the angular power spectrum of the standard weak-lensing convergence, Eq.~\eqref{eq:kappa_Phi}, is given by 
\begin{align}
    C^{\kappa_\Phi}_\ell = \frac{9}{4}\frac{H^4_0\Omega^2_m}{c^4}\int_{0}^{\chi_s}{\rm d}\chi \frac{(\chi_s-\chi)^2}{\chi^2_sa(\chi)^2}P_{\delta}\left(k=\frac{\ell}{\chi},z(\chi)\right)\,,
\label{eq:cl_kappa_Phi}
\end{align}
where $P_\delta$ is the 3D matter power spectrum. On the other hand, the angular power spectrum of the gravitomagnetic contribution, Eq.~\eqref{eq:kappa_B}, has a very similar mathematical structure to the kSZ effect --- since both effects are sourced by the rotational component of the momentum field along the LOS --- and hence we discuss these two in the next subsection.

\subsection{The convergence-kSZ cross angular power spectrum}

The gravitomagnetic contribution to the lensing convergence power spectrum is about five orders of magnitude smaller than the standard Newtonian contribution~\citep{Andrianomena:2014sya}, and even with future Stage-IV galaxy surveys such as \textsc{euclid} the former is still expected to be dominated over by cosmic variance \citep[e.g.,][]{Cuesta-Lazaro:2018uyv}. As a result, to detect the gravitomagnetic lensing effect in real observations, the lensing probe has to be cross correlated with some other observable.

As suggested by \citet{Schaefer:2005up} previously, the secondary CMB anisotropy caused by the kSZ effect (see Appendix \ref{sect:kSZ} for a brief summary) is a suitable observable to cross correlate with the gravitomagnetic lensing effect, since the former is also sourced by the integrated momentum field of matter along the LOS. 
Moreover, the kSZ effect has negligible correlation with the standard Newtonian contribution to the total lensing convergence field at the two-point level, due to the statistical isotropy of the velocity field \citep{Scannapieco:2000ut,Castro:2002df,Dore:2003ex}, which in combination with the previous point allows kSZ to single out the gravitomagnetic contribution in the lensing signal.
In other words, denoting as $b(\hat{\bf n})=-{\Delta T(\hat{\bf n})}/{\bar{T}}$ the temperature change of CMB photons along the LOS direction $\hat{\bf n}$ due to the kSZ effect, we have that the angular cross correlation between kSZ and the total convergence field (which is what observations give) reduces to
\begin{align}
   \langle b\kappa_{\rm GR}\rangle=\langle b\kappa_\B\rangle \,,
   \label{eq:kappa_GR_x_b}
\end{align}
where the angular brackets denote ensemble average.

The vanishing of the cross spectrum between $\kappa_\Phi$ and the kSZ effect can also be understood as follows: while the overdensity field can be correlated with a cluster that moves toward us, in an infinite universe there are equal chances for this to be correlated with one moving away from us, and thus the average over all possible lines of sight vanishes. At a more general level, the isotropy of the velocity field implies that, along the LOS, odd statistics of this field are subdominant with respect to even statistics \citep[][]{monin:1971v2,Jaffe:1997ye,Scannapieco:2000ut,Castro:2002df}. This feature makes kSZ an interesting candidate to potentially extract the gravitomagnetic effect in weak-lensing, and Eq.~\eqref{eq:kappa_GR_x_b} is the signal to measure the gravitomagnetic field that we will study in this paper.
In particular, in this study we will restrict our attention to the post-reionisation contribution to the kSZ signal, hence we assume that the electron density field closely follows the density field of baryons. Moreover, for simplicity we assume a fully ionised medium, i.e., we set $\chi_{\rm e}=1$ in Eq. \eqref{eq:ionised_fraction}.

The angular power spectra of the two sky observables appearing in the right hand side of Eq.~\eqref{eq:kappa_GR_x_b}, as well as their cross spectrum, can be derived as follows.
Neglecting the contribution from the longitudinal (curl-free) component of the momentum field to the kSZ effect, from Eq. \eqref{eq:kappa_B} and Eq. \eqref{eq:b_ksz} we can write the two effects as a weighted LOS integral for a general sky observable $X$ which is sourced by the rotational component of the momentum field of matter along the LOS, i.e.,
\begin{equation}\label{eq:obs_real_space}
    X(\hat{\bf n}) = \int{\rm d}\chi{K}_X(\chi)[{\bf q}_\perp\cdot\hat{\bf n}]\left(\chi\hat{\bf n},z\right)\,,
\end{equation}
where the kernels for the gravitomagnetic convergence field and the kSZ effect are respectively given by Eq.~\eqref{eq:Kernel-kappa-B}, and
\begin{equation}
    K_b(\chi) = \frac{\sigma_{\rm T}\bar{n}_{{\rm e},0}}{c}a(\chi)^{-2}e^{-\tau}\,.
    \label{eq:kernel-kSZ}
\end{equation}
As usual, the cross angular power spectrum between two observables, $X$ and $Y$, where $X,Y=b,\kappa_{\bf B}$ (and $X$ can be the same as $Y$) is defined as
\begin{equation}
    C_\ell^{XY}\delta_{\ell\ell'}\delta_{mm'} = \left\langle{a}^{X}_{\ell m}a^{Y\ast}_{\ell'm'}\right\rangle\,.
\end{equation}
After some standard derivations in the context of Limber integrals (see Appendix \ref{sect:kSZ}), it can be shown that the cross angular spectrum of these two momentum-sourced observables is given by
\begin{align}
C^{XY}_\ell = \frac{1}{2}\int{\rm d}\chi\chi^{-2}K_X(\chi)K_{Y}(\chi)P_{q_\perp}\left(k=\frac{\ell}{\chi},z(\chi)\right)\,,
\label{eq:cl_xy}
\end{align}
where $P_{q_\perp}$ is the 3D power spectrum of the rotational component of the momentum field of matter. In this case, for the power spectrum of a rotational vector field ${\bf V}$, such as $\B$ or ${\bf q}_\perp$, we use the definition
\begin{align}
\left\langle{\bf V}^i({\bf k)}{\bf V}^{*j}({\bf k}')\right\rangle=\delta^{\rm D}({\bf k}-{\bf k}')(2\pi)^3\frac{1}{2}\left(\delta^{ij}-\frac{k^ik^j}{k^2}\right)P_{\bf V}(k)\,,
\end{align}
where $\delta^{ij}$ is the Kronecker delta. Eq.~\eqref{eq:cl_xy} is the expression for both the angular auto power spectrum of $\kappa_{\bf B}$ and kSZ, and the cross angular spectrum between them. 
We remark that in the above result, the contribution from the longitudinal component of the momentum field along the LOS to the kSZ effect has been neglected. As shown by \citet{Park:2015jea}, the contribution from the longitudinal component peaks on very large angular scales, where this can dominate over the contribution from the rotational component, but it rapidly decays and becomes subdominant above $\ell\sim100$. Since we are interested in the latter regime, we expect Eq.~\eqref{eq:cl_xy} to hold up to a good approximation.
We also notice that, although the cross-correlation of either of these two observables with $\kappa_\Phi$ is expected to identically vanish due to the statistical isotropy of the velocity field \citep{Dore:2003ex}, an exact cancellation might not actually take place in observations due to, e.g., cosmic variance, which can represent a noise for the physical signal Eq.~\eqref{eq:kappa_GR_x_b}.

\section{Simulations}
\label{sec:simulations}

In order to model the convergence field Eq.~\eqref{eq:kappa_GR} and the kSZ effect Eq.~\eqref{eq:b_ksz1} we ultimately require to characterise the density and momentum field of matter, the latter being intrinsically nonlinear. While these can be respectively calculated from first- and second-order perturbation theory, at low redshift the results are expected to breakdown above $\ell\sim100$. Given that in this paper we are interested in studying the lensing-kSZ cross-correlation on smaller angular scales, and at the same time quantify the effect of cosmic variance on this signal, we therefore resort to use a suite of 30 statistically independent $N$-body simulations with $N=1024^3$ dark matter particles in a comoving box size $L_{\rm box}=1\Gpch$, which are run with the {\sc ramses} code \citep{Teyssier:2001cp}, i.e., using Newtonian gravity.
The latter is sufficient take into account the gravitomagnetic effects at leading order in the Post-Friedmann expansion, since the dynamics of CDM is not affected by $\B$ at this order, but there is an effect in the propagation of light~\citep{Bruni:2013mua}. Hence, to evaluate the gravitomagnetic convergence field in Eq.~\eqref{eq:kappa_B} we can simply use the momentum field from these Newtonian simulation.
The initial conditions were generated at $z=49$ with the {\sc 2lpt}ic code~\citep{Crocce:2006ve}, using as input a matter power spectrum from \camb~\citep{CAMB}, and the simulations are evolved from $z=49$ to $z=0$.
The cosmological parameters adopted for the simulation are $[\Omega_\Lambda$, $\Omega_m$, $\Omega_K$, $h]=[0.693,0.307,0,0.68]$ and a primordial spectrum with amplitude $A_s=2.1\times10^{-9}$, spectral index $n_s=0.96$ and a pivot scale $k_{\rm pivot}=0.05~{\rm Mpc}^{-1}$. Here $\Omega_\Lambda$ and $\Omega_K$ are, respectively, the density parameters for the cosmological constant $\Lambda$ and curvature $K$, and $h$ is the dimensionless Hubble constant, $h\equiv H_0/(100~{\rm km/s/Mpc})$.

In addition to the 30 Newtonian simulations described above, we also use a single realisation of a high-resolution, general-relativistic $N$-body simulation run with \gramses{}~\citep{Barrera-Hinojosa:2019mzo, Barrera-Hinojosa:2020arz}, which adopts the same cosmological parameters given above, and it starts from the same seed as one of the 30 Newtonian simulations.
This simulation tracks $N=1024^3$ particles in a simulation box of $L_{\rm box}=256\Mpch$, and thanks to adaptive-mesh-refinement (AMR) settings it has resolved scales down to $2\kpch$. 
Given that this simulation is fully relativistic, the gravitomagnetic field is solved and outputted by the code during the evolution, and the gravitomagnetic force acting on CDM particles is included.
This simulation has been recently used to study the vector potential of \LCDM{} in \citet{Barrera-Hinojosa:2020gnx}, where more details can be found, and it complements the suite of Newtonian simulations in two particular aspects: firstly, it serves as a fully-relativistic counterpart to the post-Friedmann approach used throughout this paper, and secondly it provides a substantial increase in resolution which can be used to test numerical resolution effects which, as we will show, can play an important role in the noise estimation for the gravitomagnetic signal from mock maps.

\section{Methodology}
\label{sec:methods}
\subsection{Modelling the observables}
\label{sec:modelling_obs}

\begin{figure*}
    \includegraphics[width=1.0\linewidth]{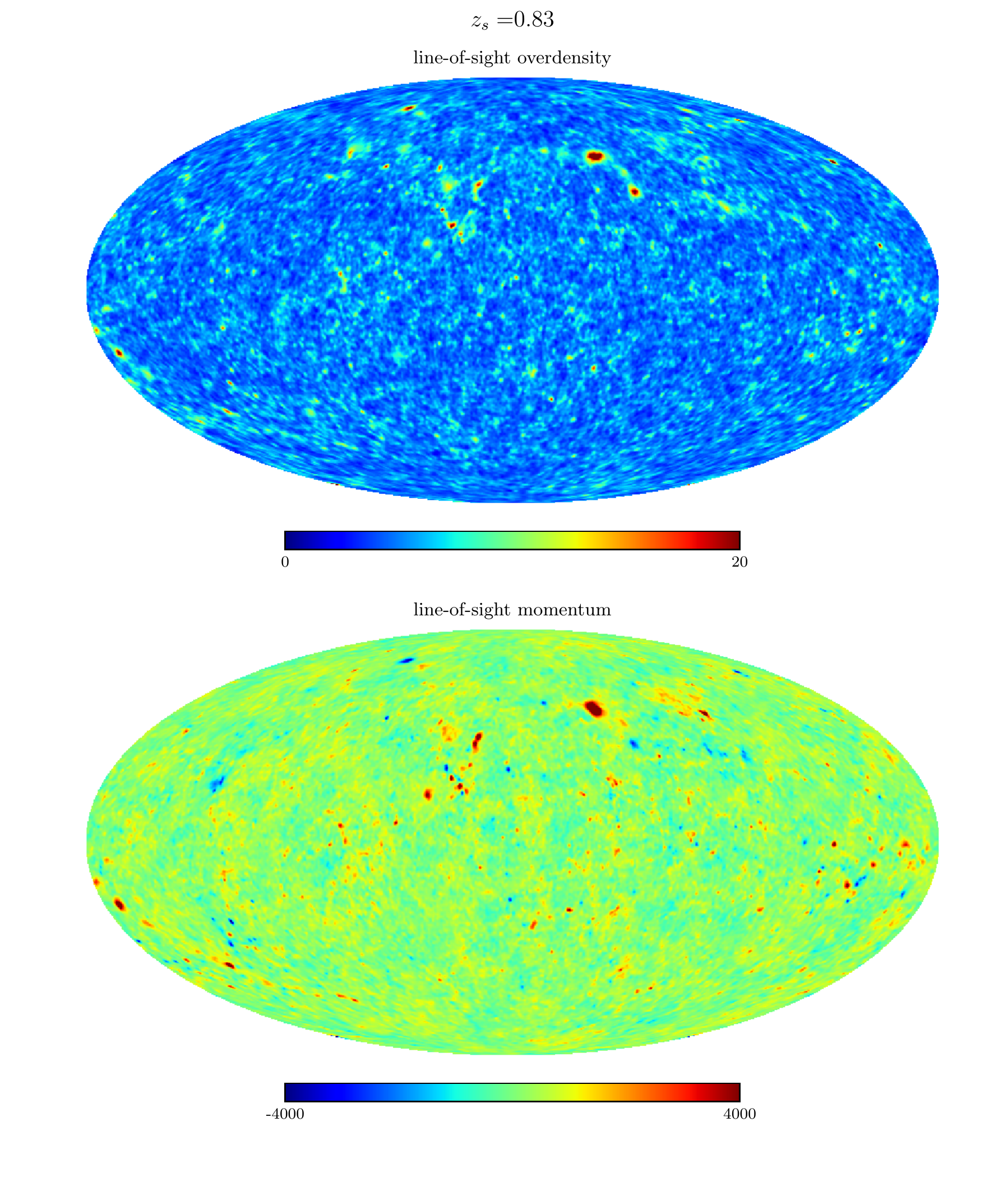}
    \caption{(Colour Online) Examples of full-sky maps for a redshift-zero observer generated by line-of-sight projections using the particle data from a $L=1\Gpch$-box simulation, up to the comoving distance $\chi_s=2\Gpch$ (corresponding to $z_s=0.83$). \textit{Top}: Projected density field. \textit{Bottom}: Projected momentum field in units of km$/$s. To help visualisation, the maps have been smoothed using a Gaussian beam with a full-width-half-maximum of 1 degree and only display a limited range of values (as indicated by the colour bars). No kernel weights have been used in these line-of-sight projections. The cross-correlation between these example maps will allow us to pick up the gravitomagnetic effect. Throughout this paper, all \textsc{healpix} maps are built using $\textsc{nside}=512$.
    }
    \label{fig:sky-maps}
\end{figure*}

To model the lensing convergence field and the kSZ effect, we take two different approaches, both of which use the particle data (positions and velocities) from the snapshots of the simulations detailed in Section \ref{sec:simulations}. In the first approach, we use this data to interpolate the density and momentum fields onto a grid, and perform LOS projections to generate mock sky maps using \href{https://healpix.sourceforge.io/}{{\sc healpix}} \citep[][]{healpix}. In the second approach, we measure the 3D power spectra of the density and momentum fields from the simulation data, and use them to evaluate the theoretical Limber integrals Eqs.~(\ref{eq:cl_kappa_Phi}, \ref{eq:cl_xy}). The 3D spectra of the density and momentum fields, which are obtained using the \dtfe{} code \citep{Schaap2000,Cautun2011} from particle data, are measured using {\sc nbodykit} \citep{nbodykit:2018}.
To single out the rotational component of the momentum field to evaluate Eq.~\eqref{eq:cl_xy}, we take the curl of this field using a 3-point finite-difference approximation, and use the identity $P_{{\bf q}_\perp}(k)=P_{\nabla\times{\bf q}}(k)/k^2$. We carry out this procedure with the 30 Newtonian $N$-body simulations.

In the case of the single \gramses{} simulation counterpart, the situation is slightly different as the vector (as well as scalar) potential values are calculated and stored by the code in the cells of the hierarchical AMR meshes, and the 3D power spectrum of the gravitomagnetic field itself is therefore measured using a code that is able to handle such mesh data directly and to write it on a regular grid by interpolation \citep{He:2015bua}. In this way, the 3D power spectrum measured from this high-resolution simulation is accurate down to $k=15\invMpch$ (see Appendix A of \citet{Barrera-Hinojosa:2020gnx}), which allow us to extend our analysis up to smaller angular scales than with the $1\Gpch$ simulations. Comparing the two sets of simulations not only allows us to assess the impact of simulation resolution, but can also serve as a cross check of the gravitomagnetic field power spectra calculated in different ways.

The sky maps in the first approach are generated using an onion-shell technique, in which an observer is placed in a random position of the simulation box and the fields are projected along different lines of sight in radial shells of fixed comoving distance, in our case with thickness of $100\Mpch{}$. Then, to generate the convergence field and kSZ maps for the given observer, the \textsc{healpix} maps of the shells (pixels) are weighted by the appropriate integration kernels corresponding to each observable, i.e., Eqs.~(\ref{eq:Kernel-lensing}, \ref{eq:Kernel-kappa-B}, \ref{eq:kernel-kSZ}). In this process, the simulation box is tiled multiple times to cover the volume enclosed by a sphere up to the comoving distance of the source, $\chi_s$, if needed. 
In order to avoid noticeable effects due to the tiling of the $1\Gpch$ box in the sky maps --- which can introduce repeated structures --- we have chosen $\chi_s=2000\invMpch$ ($z_s=0.83$) as the maximum comoving distance in the LOS projections.
Finally, the angular power spectra of a resulting sky map, or the cross correlation between two different maps, is measured with {\sc anafast} subroutines included in {\sc healpix}.

Figure \ref{fig:sky-maps} is an example of the full-sky maps of projected density and momentum along the LOS up to $z_s=0.83$ for an observer located at $z=0$, without applying any kernel weights, using the data from one of the $1\Gpch$-box simulations.
A visual inspection of Fig.~\ref{fig:sky-maps} shows a clear correlation between the density (top) and momentum (bottom) field maps. Along the overdense lines of sight, the projected momentum field can be in either of the two directions; in an infinite universe one would expect this to be an exact symmetry, which makes the cross correlation between the projected density and momentum fields vanish. However, due to the finite box size, this does not happen exactly, as is evident from the fact that in this particular case there are more lines of sight with positive values of the projected momentum field (i.e. pointing away from the observer) than negative values. This means that the kSZ-$\kappa_\Phi$ cross correlation measured from sky maps made from these simulations would not exactly vanish on large scales, although it is expected to eventually vanish on small scales, where the effect of cosmic variance is suppressed. It is therefore important to accurately quantify the degree of cross-correlation between these two effects, which can become a source of noise for the physical signal in Eq.~\eqref{eq:kappa_GR_x_b}.
A standard way to measure the degree of correlation between two random fields is by the cross-correlation coefficient, defined as
\begin{equation}
    {\rm Corr}_{XY}(\ell)=\frac{C^{XY}_\ell}{\sqrt{C^{XX}_\ell C^{YY}_\ell}}\,.
\label{eq:corr-c-xy}
\end{equation}

We will proceed to measure the auto- and cross-angular power spectra from the maps and compare with the theoretical predictions in Section~\ref{subsect:noise}. In particular, we will present the cross-correlation coefficient between $\kappa_\Phi$ and the kSZ effect measured from the sky maps, and show that the measurement depends sensitively on simulation resolution. However, provided that the simulation resolution is high enough, our result suggests that the covariance of the kSZ-$\kappa_\Phi$ cross angular power spectrum found in the map measurements agrees well with the corresponding theoretical prediction of sample variance,  Eq.~\eqref{eq:std-general-theory} below.

\begin{figure*}
    \centering
    \includegraphics[width=0.9\linewidth]{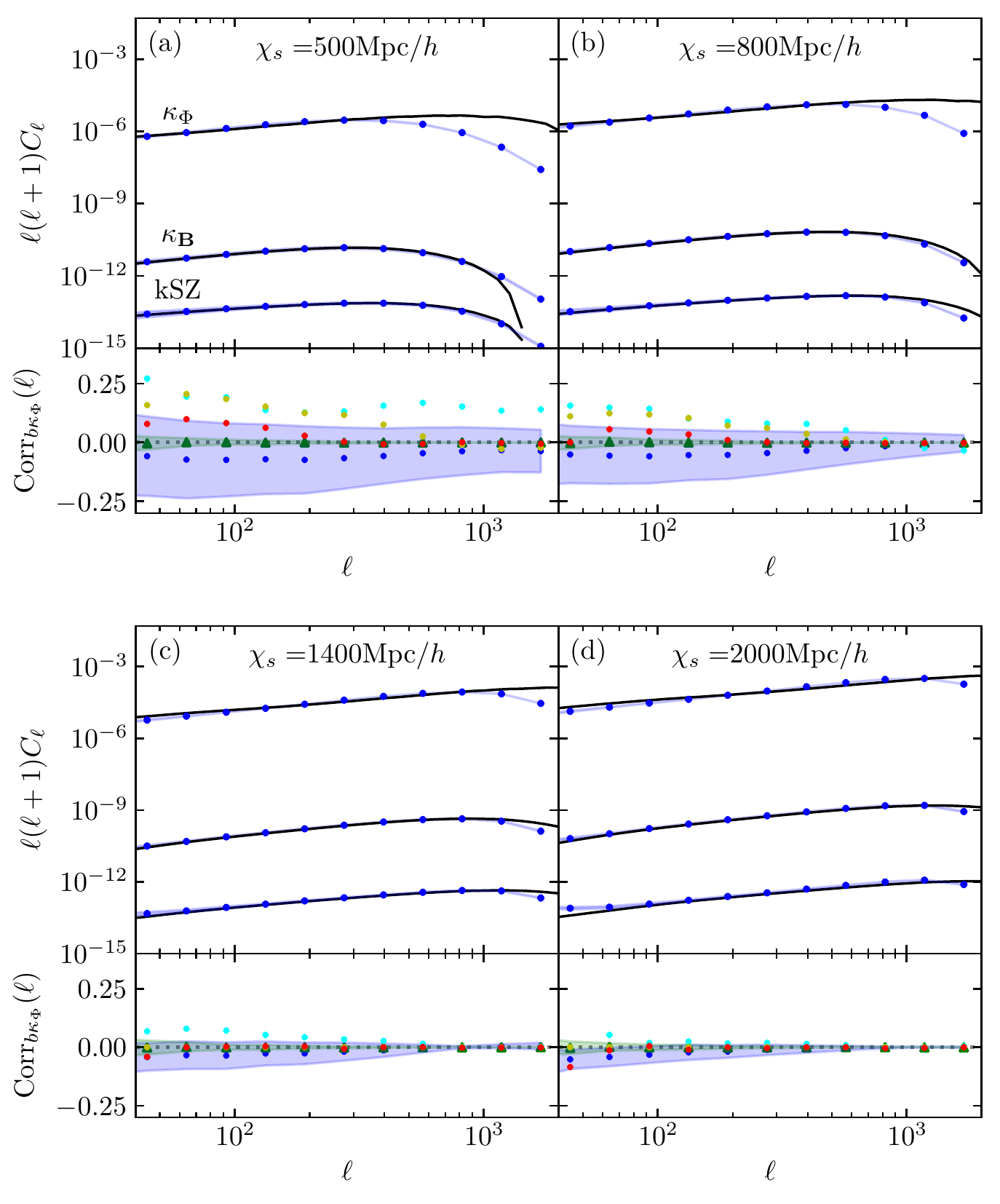}
    \caption{(Colour Online) \textit{Top subplots in panels (a)--(d)}: Angular power spectra of the Newtonian ($\kappa_\Phi$) and gravitomagnetic ($\kappa_\B$) contributions to the convergence field and the kSZ effect ($b$) for different comoving distances up to $\chi_s=2000\Mpch$ (which corresponds to $z_s=0.83$) from a redshift-zero observer. 
    The black solid lines show the Limber-approximated integrals evaluated with the 3D power spectra measured from the $1\Gpch$-box simulations, while blue circles show the mean of the 30 {\sc healpix} maps from the same simulations, and shaded blue region the corresponding $1\sigma$ standard deviation.  
    \textit{Bottom subplots in panels (a)--(d)}: Cross-correlation coefficient of kSZ-$\kappa_\Phi$ {(expected to be zero in theory)}. The blue circles and blue shaded area respectively correspond to the mean and $1\sigma$ scatter of the `same-seed' case, while green circles represent the analogue result for the `cross-seed' case (see main text for details). The cyan, yellow and red circles correspond to the cross-correlation coefficient obtained from a single, same realisation (seed), with box size $L=1\Gpch$, $L=500\Mpch$, and $L=256\Mpch$, respectively. The results use 11 $\ell$-bins spaced logarithmically between $\ell_{\rm min}=40$ and $\ell_{\rm max}=2000$. {The kSZ-$\kappa_\Phi$ cross-correlation is only fully consistent with zero for the simulation with the highest resolution ($L=256\Mpch$), indicating that it is necessary to use this high resolution simulation to make reliable predictions for the signal and noise of the gravitomagnetic effect.} 
    }
    \label{fig:kappa_Lbox_test}
\end{figure*}

\subsection{Comparison of auto- and cross-power spectra from mock maps and the Limber approximation}
\label{subsect:noise}

Given that the cross correlation between $\kappa_\Phi$ and the kSZ effect vanishes theoretically, the resulting data when measuring this quantity from the sky maps can be very noisy, with strong fluctuations around 0 from one $\ell$-mode to the next. Thus, for the subsequent analysis we will bin the spectrum data in $\ell$-space, which will cancel out most of the oscillations and reduce the noise, hence leading to smoother measurements. In the remainder of this Section, we bin the data into 11 $\ell$-bins spaced logarithmically between $\ell_{\rm min}=40$ and $\ell_{\rm max}=2000$.

The top panels in the two rows of Fig.~\ref{fig:kappa_Lbox_test} show the angular power spectra of the two contributions to the lensing convergence field, i.e., $\kappa_\Phi$ (top curves) and $\kappa_\B$ (bottom curves) in Eq.~\eqref{eq:kappa_GR}, and of the kSZ effect ($b$; middle curves), at different comoving distances of the lensing source, $\chi_s$ (different panels). These spectra are measured from the {\sc healpix} maps (circles) or calculated with the Limber-approximation integrals (black solid line), and for the former we show the mean and $1\sigma$ regions from the 30 $1\Gpch$ realisations. We find an overall very good agreement between the two methods, especially for $\kappa_\B$ and kSZ, up to $\ell\sim1000$, where the pixel resolution effect in the \textsc{healpix} maps starts to appear. Notice that the ratio between the angular power spectrum of $\kappa_\B$ and $\kappa_\Phi$ is about $\mathcal{O}(10^{-5})$, which is of the same order as the ratio between the 3D power spectra of the corresponding potentials \citep[see, e.g., Fig.~5 of][]{Barrera-Hinojosa:2020gnx}.

The bottom panels in the two rows of Fig.~\ref{fig:kappa_Lbox_test} show the cross-correlation coefficient, Eq.~\eqref{eq:corr-c-xy}, for kSZ-$\kappa_\Phi$ measured from the mock maps. To study this cross correlation in detail, we measure it in two ways: first by picking both sky maps from the same realisation\footnote{We always use a single simulation box, if necessary tiling it multiple times as described above, to obtain a given sky map. In this sense, each realisation of map corresponds to a single realisation of simulation.} (dubbed `same seed' below), and then picking each map from a different realisation (dubbed `cross seed'). In the first case, we then average over all 30 realisations as before, while in the second case we average over all possible combinations. We find that these approaches give very different results. First, we note that in the `cross seed' case, the mean (green triangles) is consistent with zero, the standard deviation (green shaded region) is symmetric around the horizontal dashed line ($0$), and its magnitude consistently decreases toward small angular scales. 
In contrast, in the same-seed case we find that on large and intermediate scales the mean (blue circles) lies beyond the $1\sigma$-region of the cross-seed case, 
and the standard deviation (blue shaded region) is much larger than in the previous case and does not consistently decrease with $\ell$.

In order to pinpoint the origin of the above discrepancy, we have conducted a test of the numerical resolution. In addition to the `same seed' and `cross seed' results, in the bottom panels of the two rows in Fig.~\ref{fig:kappa_Lbox_test} we show the cross-correlation coefficients measured from sky maps made from simulations that use a single, fixed, initial condition random seed, with three different box sizes: $1h^{-1}$Gpc (cyan symbols), $500\Mpch$ (yellow) and $256h^{-1}$Mpc (red). We find that: ($i$) the cross-correlation coefficient consistently decreases with increasing resolution, with the $256h^{-1}$Mpc box giving cross-correlation coefficient values that are very close to 0; ($ii$) the deviation of ${\rm Corr}_{b\kappa_\Phi}(\ell)$ from 0 is strongest for lower lensing source reshift (smaller $\chi_s$), which is likely because these maps enclose a much smaller volume. Then, since the LOS projection (in terms of the Limber approximation) probes modes $k=\ell/\chi$, these sky maps critically depend on contributions from small scales which may not be well resolved by some of the simulations.
Nevertheless, even for $\chi_s=500h^{-1}$Mpc, the $256h^{-1}$Mpc box gives a ${\rm Corr}_{b\kappa_\Phi}$ that is very close to $0$. Hence, we conclude that this is a numerical resolution effect due to unresolved scales close to the observer's location as a consequence of lack of resolution in the simulations.

\begin{figure}
    \centering
    \includegraphics[width=1.0\linewidth]{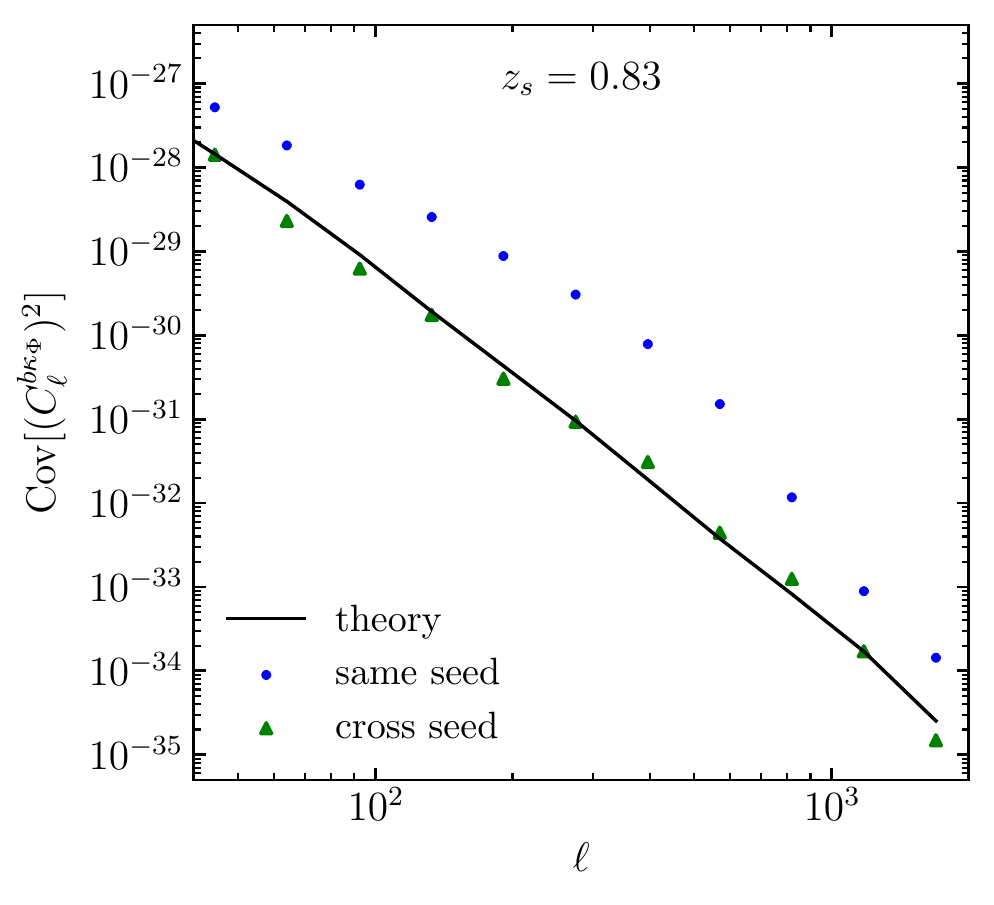}
    \caption{(Colour Online) Comparison of the covariance of the kSZ ($b$)-$\kappa_\Phi$ cross-angular power spectrum from {\sc healpix} maps and the theoretical prediction, Eq. \eqref{eq:std-general-theory}, at $z_s=0.83$.
    The results use 11 $\ell$-bins spaced logarithmically between $\ell_{\rm min}=40$ and $\ell_{\rm max}=2000$. {The `cross seed' case agrees with the theoretical prediction very well, indicating the robustness for the estimate of cosmic variance.}   
    }
    \label{fig:std_cross_validation}
\end{figure}

We now compare the covariance of the kSZ-$\kappa_\Phi$ cross correlation measured from the sky maps with the theoretical expectation for the effect of cosmic variance. In the latter case, the covariance of the cross angular power spectrum between two Gaussian fields, $A$ and $B$, is given by \citep[e.g.,][]{Cabre:2007rv}
\begin{align}
    {\rm Cov}\left[\left(C^{AB}_\ell\right)^2\right] = \frac{1}{\Delta\ell f_{\rm sky}(2\ell+1)}\left[\left(C^{AB}_\ell\right)^2+C^A_\ell C^B_\ell\right]\,,
\label{eq:std-general-theory}
\end{align}
where $f_{\rm sky}$ is the observed fraction of the sky, and $\Delta\ell$ is the width of the multipole bins, which is assumed to be independent of $\ell$, i.e., no off-diagonal terms in the covariance matrix. 
For the particular case of $\kappa_\Phi$ and kSZ, the theoretical cross angular power spectrum in the right hand side of Eq.~\eqref{eq:std-general-theory} vanishes and only the second term in the square bracket contributes.

Figure \ref{fig:std_cross_validation} shows the covariance of the cross angular spectrum between $\kappa_\Phi$ and the kSZ effect at $z_s=0.83$ measured from the same-seed maps (blue circles) and cross-seed maps (green triangles) from the $1\Gpch{}$-box simulations, and the theoretical prediction  Eq.~\eqref{eq:std-general-theory} (solid black line). We find that the latter is in very good agreement with the results from the cross-seeds case maps across all scales, while the covariance in the same-seed case maps, which is affected by the resolution effects discussed above, can become over one order of magnitude larger at around $\ell\simeq500$ and thus strongly degrade the signal-to-noise estimation of the gravitomagnetic signal Eq.~\eqref{eq:kappa_GR_x_b}. As in the above discrepancy, this is not unexpected since the sky observables take the form of LOS integrals, hence the numerical resolution errors due to unresolved scales close to the observer's location can propagate up to higher redshifts (comoving distances) and contaminate the final result. It is worthwhile to remark that, even though the lensing kernel down-weights the radial shells that are closer to the observer, and hence suppresses the relative contribution of these numerical resolution effects when projecting up to a high redshift (e.g., $z_s=0.83$ as in our case), the result shows that the cross correlation is still considerably large compared to the effect of sample variance only.

\section{Results}
\label{sec:results}

\begin{figure*}
    \centering
    \includegraphics[width=1.0\linewidth]{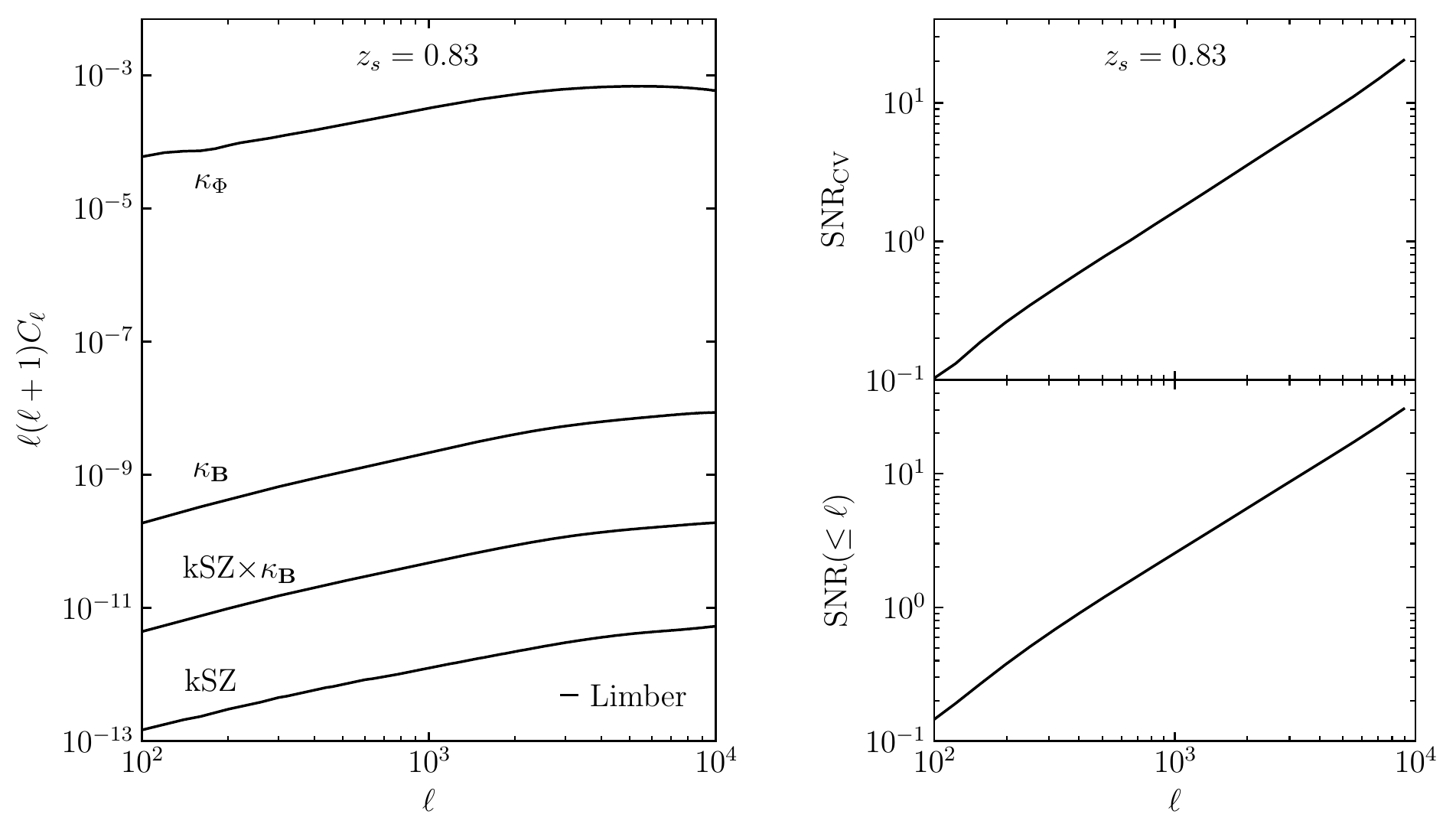}
    \caption{\textit{Left panel}: Angular power spectrum of the two contributions to the lensing convergence field and the kSZ effect ($b$), and the cross angular power spectrum of kSZ-$\kappa_\B$. These are obtained from Limber-approximated integrals evaluated with the 3D power spectra measured from the simulation with box size $L=256\Mpch$.
    \textit{Top right panel}: Theoretical signal-to-noise ratio (SNR) for the kSZ-$\kappa_\B$ cross-correlation, corresponding to the idealised case where the noise is dominated by the cosmic variance of the kSZ-$\kappa_{\rm GR}$ signal itself.
    \textit{Bottom right panel}: Cumulative SNR corresponding to the top plot.
    The results use 23 $\ell$-bins spaced logarithmically between $\ell_{\rm min}=40$ and $\ell_{\rm max}=10^4$.    
    }
    \label{fig:C_l-SNR}
\end{figure*}

In this section, we will more quantitatively assess the gravitomagnetic lensing effect signal and its detectability. This will be done in the context of cross correlating two observables --- a total lensing convergence field containing the gravitomagnetic effect, and a total CMB temperature map that contains the (integrated) kSZ effect. We quantify this using the usual signal-to-noise ratio (SNR),
\begin{equation}\label{eq:SNR_general}
    \left(\frac{S}{N}\right)^2_\ell = \frac{\left(C^{AB}_\ell\right)^2}{{\rm Cov}\left[\left(C^{AB}_\ell\right)^2\right]}\,,
\end{equation}
where $C^{AB}_{\ell}$ is the cross angular spectrum between two generic observables $A$ and $B$, and ${\rm Cov}$ denotes the covariance matrix. In the ideal scenario, the covariance matrix in Eq. \eqref{eq:SNR_general} is dominated by the effect of cosmic variance and Eq.~\eqref{eq:std-general-theory} directly applies. However, for a realistic estimation of the SNR, the covariance also needs to include the following two contributions: $(i)$ instrumental noises in the sky maps of $A$ and $B$; $(ii)$ spurious signals caused by other physical effects, such as the primary and other secondary CMB anisotropies in the case of a CMB temperature map. 

Our main objective is to forecast the detectability of the gravitomagnetic effect for various future galaxy surveys and CMB experiments (Section~\ref{sec:detectability}). However, before that, we will first calculate a `theoretical SNR' (Section~\ref{sec:theoretical_SNR}), by applying Eq.~\eqref{eq:SNR_general} while neglecting all instrumental noises and considering a pure kSZ map with no other CMB primary or secondary effects. The latter is useful for assessing, in an idealised situation, the potential of isolating the gravitomagnetic contribution to the total lensing signal by cross-correlating with kSZ --- this can serve as an upper bound of the SNR in real observations.

\subsection{Theoretical signal-to-noise ratio}
\label{sec:theoretical_SNR}

Let us first investigate the SNR for the kSZ-$\kappa_\B$ cross correlation in the most idealised case, i.e., accounting for only the variance contributed by $\kappa_{\rm GR}$ and kSZ ($b$) themselves. We will include other source of noise such as the primordial CMB and instrumental noise in Section \ref{sec:detectability}.

Because of the good agreement in the noise predictions from theory and maps shown in Fig.~\ref{fig:std_cross_validation}, to calculate the SNR we resort to using the Limber prediction Eq.~\eqref{eq:cl_xy} to model the signal, taking as input the 3D power spectrum measured from the high-resolution simulation, and use Eq.~\eqref{eq:std-general-theory} to quantify the noise, with the two fields $A, B$ being respectively the kSZ contribution to the CMB temperature fluctuation, $b$, and the total lensing convergence, $\kappa_{\rm GR}=\kappa_\Phi+\kappa_{\bf B}$.
The angular power spectra are binned into 23 $\ell$-bins logarithmically spaced between $\ell_{\rm min}=40$ and $\ell_{\rm max}=10^4$. Then, using the fact that, at the theory level, the kSZ-$\kappa_\Phi$ cross correlation vanishes and hence does not contribute to the first noise term in the square bracket of Eq.~\eqref{eq:std-general-theory}, the SNR becomes
\begin{align}\label{eq:SNR-theory}
   \left( \frac{S}{N}\right)_{\ell,{\rm CV}}
    &=\sqrt{\Delta\ell f_{\rm sky}(2\ell+1)}\frac{C^{b\kappa_\B}_\ell}{\sqrt{\left(C^{b\kappa_\B}_\ell\right)^2 +C^b_\ell C^{\kappa_{\rm GR}}_\ell}} \nonumber \\
    &\approx\sqrt{\Delta\ell f_{\rm sky}(2\ell+1)}\frac{C^{b\kappa_\B}_\ell}{\sqrt{C^b_\ell C^{\kappa_\Phi}_\ell}}\,,
\end{align}
where in the second line we have approximated $C^{\kappa_{\rm GR}}_\ell\approx C^{\kappa_{\Phi}}_\ell$ since, as shown in Fig.~\ref{fig:kappa_Lbox_test} and Fig.~\ref{fig:C_l-SNR}, $C^{\kappa_{\B}}_\ell$ is suppressed by about five orders of magnitude with respect to the Newtonian contribution, and we have also used that $\left(C^{b\kappa_\B}_\ell\right)^2\ll C^b_\ell C^{\kappa_\Phi}_\ell$ (as shown by the left panel of Fig.~\ref{fig:C_l-SNR}) to neglect the first term in the denominator. Notice that we have used the subscript ${\rm CV}$ to highlight that, to obtain this theoretical SNR, only the cosmic variances in $\kappa_{\rm GR}$ and kSZ ($b$) are included in the noise.

Given that the high-resolution simulation is fully relativistic, instead of evaluating Eq.~\eqref{eq:cl_xy} using the 3D power spectrum of the momentum field measured from the simulation, in this case we directly use the 3D power spectrum of the gravitomagnetic field that is calculated and outputted by 
\textsc{gramses} during the simulation, $P_{\B}(k)$, and the integration kernel for the convergence field Eq. \eqref{eq:kappa_B} is modified according to Eq.~\eqref{eq:momentum-constraint}\footnote{Although, rigorously speaking, \gramses{} uses a different gauge than the $N$-body gauge used by the Newtonian simulations~\citep{Fidler:2016tir}, both share the same definition of spatial coordinates, and the gravitomagnetic potential indeed corresponds to the gauge-invariant one defined in the Poisson gauge.}. Conversely to the logic behind the Post-Friedmann (or Post-Newtonian) approach --- in which the gravitomagnetic effect is ultimately written in terms of the rotational modes of the 3D momentum field --- in this case we use Eq.~\eqref{eq:momentum-constraint} to convert $P_\B(k)$ into $P_{{\bf q}_\perp}(k)$ to evaluate the kSZ effect using the same spectrum data.  At this point, it is worthwhile to remark that the gravitomagnetic potential power spectrum measured from the high-resolution simulation (and correspondingly the 3D momentum power spectrum) suffers from a power suppression due to the small box size \citep{Zhang:2003nr,Iliev:2006un}. Indeed, it has been found that this effect appears prominently if the matter-radiation equality scale is not sampled \citep{Adamek:2015eda,Barrera-Hinojosa:2020gnx}. As discussed in the Appendix B of \citet{Park:2013mv}, in the context of the momentum power spectrum and the kSZ effect (which formally involves the same calculation), the large-scale power loss can be corrected for by using perturbation theory. For this, we calculate the ratio between the second-order perturbation theory predictions of $P_{\B}(k)$, Eq. \eqref{eq:B_spectrum}, evaluated in two ways: one which matches the simulation results on large scales (i.e., which is also suppressed by a large-scale cut-off scale\footnote{This is achieved by restricting the $k$ range (in particularly the lower end) for the matter and velocity divergence power spectra used in the evaluation of the perturbation-theory result, Eq. \eqref{eq:B_spectrum}, to the same as probed by the high-resolution simulation.}), and another which does not include any cut-off and hence does not miss any power on large scales. Then, to get the corrected power spectrum we multiply this ratio to the $P_{\bf B}(k)$ measured from the simulation, and use this to evaluate Eq.~\eqref{eq:cl_xy}. Although we repeat this procedure for each available snapshot, we have checked that this correction factor is redshift-independent.

Another important aspect to take into account when evaluating the Limber integrals is the time evolution of the 3D spectra. Given that we can only measure these from a finite number of snapshots, to parameterise the time evolution of $P_\B(k,z)$ and $P_\Phi(k,z)$ we measure these from the available simulation snapshots ($z=0,0.5,1,1.5$) and interpolate among them. For the $\kappa_\Phi$ case we use the linear growth rate $D_+$, given by \citep{Linder-growth}
\begin{equation}
D_+(a)=\exp{\int_1^a {\rm d}\ln{a}'\Omega_m(a')^{6/11}}\,,
\end{equation}
with $\Omega_m(a)=\Omega_{m}a^{-3}/(H/H_0)^2$, as the `time' variable for the interpolation; more explicitly, the interpolation is linear in $D_+^2$. 
Since $P_\B(k,z)$ is sourced by the rotational component of the momentum field ${\bf q}=(1+\delta){\bf v}$, to interpolate this for the calculation of kSZ and $\kappa_\B$ we also use the linear continuity equation,
\begin{equation}
   {\bf v}({\bf k},a)=-i\frac{Hf}{k^2}{\bf k}{\delta}({\bf k},a)\,, \label{eq:continuity-fourier-space-sol}
\end{equation}
where $f={\rm d}\ln{D_+}/{\rm d} \ln{a}$ is the linear growth rate; here $\left(HfD_+\right)^2$ is used as the `time' variable for the interpolation to ensure that it gives the correct time-evolution behaviour at large linear scales. Evidently, these interpolations involve a certain degree of approximation at the small nonlinear scales, but we have checked that our result does not change significantly if we use fixed simulation snapshots or different time interpolation schemes.

Figure \ref{fig:C_l-SNR} represents one of the main results of this paper. The left panel shows the angular power spectra of the different effects based on the high-resolution simulation, which allows to resolve scales down to $\ell=10^4$. The top right panel of Fig.~\ref{fig:C_l-SNR} shows the theoretical SNR, in which the error is calculated using  Eq.~\eqref{eq:SNR-theory}, i.e., by only including the effects of sample variances in the kSZ-$\kappa_{\rm GR}$ signal, with the angular power spectra therein corresponding to those shown in the left panel. We find that, with $z_s=0.83$, a SNR of $\simeq10$ is achieved at $\ell\simeq5000$, while this can reach $\simeq20$ at $\ell\simeq10^4$. The bottom right panel of Fig.~\ref{fig:C_l-SNR} shows the cumulative SNR corresponding to the top panel of the same figure, which can reach almost 15 (30) at $\ell\simeq5000$ ($10^4$). These estimates will, of course, be downgraded once we have included realistic instrument noises and other spurious effects, as discussed in the next subsection. The same is expected to occur when baryonic effects are taken into account, although the latter is beyond the scope of this paper.

\subsection{Detectability with current and future observations}
\label{sec:detectability}

\begin{table}
	\centering
	\caption{Experimental specifications for the weak lensing surveys and CMB experiments considered in this work.}
	\label{tab:lensing-surveys-specs}
	\begin{tabular}{lccr} 
		\hline
		\hline
		Survey & $n_g$ (galaxies per arcmin${}^2$) & $\sigma_\epsilon$ & $f_{\rm sky}$ \\
		\hline
		\textsc{euclid} & 30 & 0.22 & 0.36 \\
		\textsc{lsst} & 40 & 0.22 & 0.5 \\
		\hline
		\hline
		Experiment   & $\theta_{\rm FWHM}$ [arcmin] & $\Delta_T$ [$\mu$K-arcmin] & $f_{\rm sky}$ \\
		\hline
		\textsc{planck} & 5 & 3.1 & 0.82 \\		
		\textsc{cmb}-\textsc{s4} & 1.4 & 1 & 0.4 \\		
		Simons Obs. & 1.4 & 6 & 0.4 \\		
		\hline
		\hline
	\end{tabular}
\end{table}

\begin{figure*}
    \centering
    \includegraphics[width=1.0\linewidth]{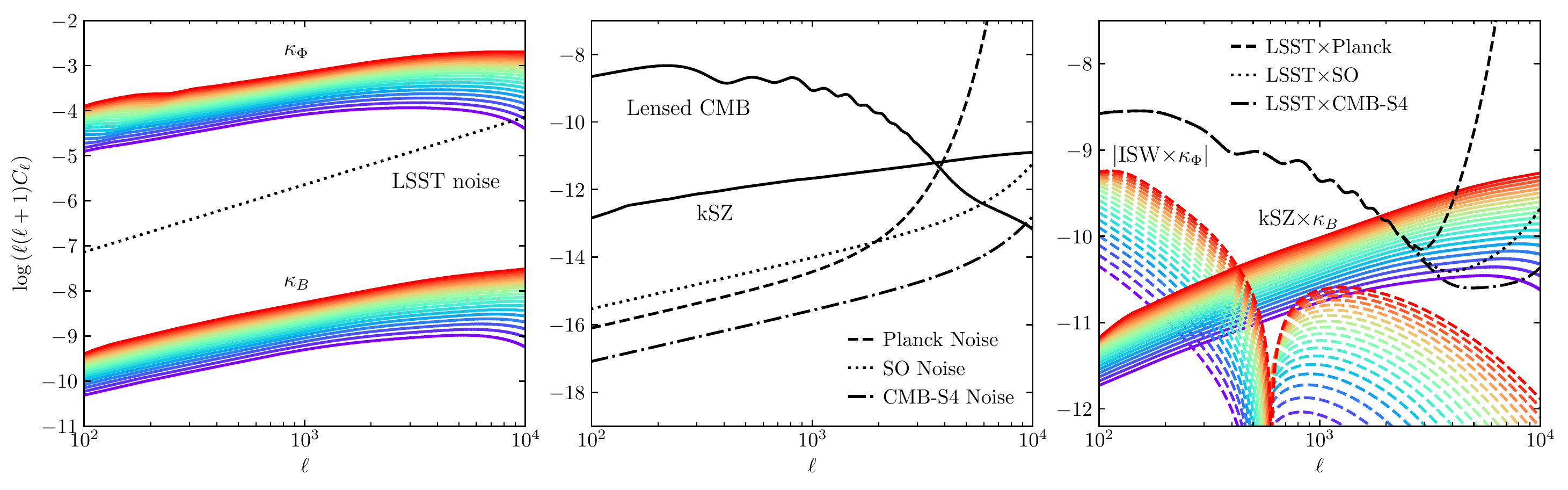}
    \caption{(Colour Online) Comparison of the various angular power spectra and noise levels of weak-lensing surveys and CMB experiments. \textit{Left panel}: Newtonian (upper solid lines) and gravitomagnetic (lower solid lines) contributions to the lensing convergence field, which indicate that the latter is around 5 orders of magnitude smaller, and is well below the expected noise level of future weak lensing surveys such as \textsc{lsst}. \textit{Middle panel}: kSZ effect and the lensed CMB signals (as indicated by legends), and the noise levels for three CMB experiments: \textsc{planck} (dashed), \textsc{so} (dotted) and \textsc{cmb-s4} (dot-dashed). The kSZ effect dominates over the lensed CMB signal at $\ell\gtrsim3500$. \textit{Right panel}: the cross spectrum of kSZ-$\kappa_\B$ (solid lines), which is the signal we are after, and the absolute value of the cross spectrum between the ISW effect and $\kappa_\Phi$ (dashed lines), which represents a potential source of contamination for the kSZ-$\kappa_\B$ signal. The dashed, dotted and dot-dashed black lines are, respectively, the expected noise level for the cross correlation of weak lensing data from an \textsc{lsst}-like survey
    and CMB data from \textsc{planck}, \textsc{so} and \textsc{cmb-s4}.
    The kSZ-$\kappa_\B$ signal is well above the noise levels of future experiments on scales $\ell\gtrsim3000$.
    In all panels, the different colours correspond to lensing source redshifts between $z_s=0.4$ (purple, lowest amplitude) and $z_s=1.4$ (red, highest amplitude), with a separation $\Delta z_s$ equivalent to a comoving distance of $\chi=100\Mpch$.
    }
    \label{fig:C_l_LSST_x_CMB_noise}
\end{figure*}

Let us now investigate the detectability of the gravitomagnetic signal with current and future observations.
In real observations, the kSZ effect is imprinted in the measured CMB temperature map along with a number of primary and secondary anisotropies. Because the latter is what will be used to cross correlate with weak lensing, to assess the detection of the kSZ-$\kappa_\B$ cross-correlation, we need to consider all the relevant contributions contained in a full CMB map. In particular, it is essential to include the cosmic variance effect from the primordial CMB, as this signal dominates over kSZ on scales down to $\ell\sim3000$. We will discuss these effects and how they are expected to affect the sought-after physical signal below.

The signal-to-noise per individual mode of the lensing-kSZ cross correlation, Eq.~\eqref{eq:kappa_GR_x_b}, is given by Eq.~\eqref{eq:SNR_general}, which can now be written more explicitly as
\begin{align}
    \left(\frac{S}{N}\right)^2_\ell&=\frac{\left(C^{b\kappa_{\bf B}}_\ell\right)^2}{{\rm Cov}\left[\left(C^{\mathrm{T}\kappa_{\rm GR}}_\ell\right)^2\right]} \,,
\end{align}
where $C^{b\kappa_{\bf B}}_\ell=C^{b\kappa_{\rm GR}}_\ell$ is again the physical signal we are after, while $C^{\mathrm{T}\kappa_{\rm GR}}$ is the cross angular power spectrum between the total CMB temperature map (${\rm T}$) and the total lensing convergence field, $\kappa_{\rm GR}$. Neglecting the correlations induced by the incomplete sky coverage, the covariance matrix can be approximated as
\begin{align}\label{eq:obs_cov}
  \frac{{\rm Cov}\left[\left(C^{\mathrm{T}\kappa_{\rm GR}}_\ell\right)^2\right]}{\left[\Delta\ell f_{\rm sky}(2\ell+1)\right]^{-1}} \approx {\left(C^{b\kappa_{\bf B}}_\ell\right)^2+\left(C^{\rm T}_\ell + N^{\rm T}_\ell\right)\left(C^{\kappa_{\rm GR}}_\ell + N^{\kappa_{\rm GR}}_\ell\right)}\,, 
\end{align}
where $C^{{\rm T}}_\ell$ is the total angular power spectrum of the CMB temperature, which includes the kSZ effect, the integrated Sachs-Wolfe (ISW) effect, and the weak lensing of the CMB. Frequency-dependent secondary effects on the CMB, such as bright radio galaxies, the cosmic infrared background (CIB) and thermal SZ (tSZ) effect are expected to dominate over the signal we are after at $\ell$ of a few thousands to ten thousands \citep[e.g.][]{ACT2020,SPTpol2021}. In principle, these can be removed with multi-frequency observations. However, due to the possible imperfect modelling for their spectra and limited range of frequency coverage, residuals contaminations are likely to remain in the foreground reduced CMB temperature maps. These can be dealt with further by modelling their clustering with free parameters \citep[e.g.][]{PlanckCIB2014,ACT2020,SPTpol2021}. In addiction, due to the unique dipole feature of the gravitomagnetic effect imprinted on both the kSZ and lensing signal, some matched filter techniques should be effective for filtering out the signal from other contaminations (a detailed investigation into this is beyond the scope of the present paper).

For the lensed CMB angular-power spectrum we use the output from {\sc camb}, to which we add the kSZ contribution calculated using the Limber approximation, Eq. \eqref{eq:cl_xy}. Since the available simulation data only covers up to $z=1.5$, kSZ is only integrated up to this redshift (rather than up to $z\sim6$, which corresponds to the end of reionisation). 
When cross correlating a CMB map including other secondary effects and a galaxy weak lensing map, we need to consider if these secondary CMB signals can lead to spurious correlations which contaminate the sought-after signal, $C^{b\kappa_{\bf B}}$, particularly through cross correlations with $\kappa_\Phi$, because $|\kappa_\Phi|\gg|\kappa_{\bf B}|$, so that any such spurious signal can potentially be as strong as, if not stronger than, $C^{b\kappa_{\bf B}}$ itself. 
At small angular scales, the CMB power spectrum is dominated by lensing, with the lensed temperature at sky position $\vec{\theta}$ approximately given by
\begin{equation}\label{eq:lensed_CMB_T}
T^{\rm lensed}(\vec{\theta}) = T^{\rm unlensed}(\vec{\theta}) + \vec{\nabla}T\cdot\vec{\nabla}\phi\,,
\end{equation}
where $\phi$ is the CMB lensing potential. Because $\vec{\nabla}{T}$ has no correlation with the late-time large-scale structures in theory, we expect the correction term in Eq.~\eqref{eq:lensed_CMB_T} to have zero theoretical cross correlation with weak lensing $\kappa_\Phi$: note this is different from the cases of cross correlating the CMB lensing deflection angle or convergence field (in both cases $\vec{\nabla}T$ has been removed through de-lensing reconstruction \citep{Plancklensing}), or the squared CMB field \citep[e.g.,][]{Dore:2003ex}, with weak lensing. 
On the other hand, the ISW effect, along with its nonlinear counterpart, the Rees-Sciama (RS) effect, can have a nonzero cross correlation with weak lensing \citep{Hu:2001fb}; we have explicitly calculated this spurious signal using the method described in Appendix \ref{sect:isw-wl-x}, and found it to be subdominant compared to the kSZ-$\kappa_{\mathbf{B}}$ cross power spectrum $C^{b\kappa_{\mathbf{B}}}_\ell$ at the small angular scales of interest to us, as will be discussed below. Therefore, in Eq.~\eqref{eq:obs_cov} we have neglected the contribution from $C_{\ell}^{\mathrm{ISW}\kappa_\Phi}$, thus approximating $\left(C^{{\rm T}\kappa_{\rm GR}}_\ell\right)^2$ by $\left(C^{b\kappa_{\B}}_\ell\right)^2$.

In Eq.~\eqref{eq:obs_cov}, $N^X_\ell$ represents the contribution from the instrumental noise to the measured angular power spectrum of each effect. For the lensing signal (cosmic shear), we assume that the dominant error comes from the intrinsic ellipticity of galaxies, i.e.,
\begin{equation}
    N^{\kappa_{\rm GR}}_\ell=\frac{\sigma^2_\epsilon}{n_{\rm gal}}\,,
\end{equation}
where $\sigma^2_\epsilon$ is the variance of the intrinsic ellipticity of galaxies, and $n_{\rm gal}$ the number of source galaxies per arcmin${}^2$.
For the CMB signal, we consider the error due to instrumental noise and beam smearing, given as \citep{Knox:1995dq}
\begin{equation}
    N^{\rm T}_\ell=\left(\frac{\Delta_T}{\bar{T}}\right)^2\exp{\left[\ell^2\theta^2_{\rm FWHM}/(8\ln{2})\right]}\,,
\end{equation}
where $\Delta_T$ is the noise level, $\bar{T}$ is the mean temperature of the CMB, and $\theta_{\rm FWHM}$ is the full width at half-maximum of the beam.
Table \ref{tab:lensing-surveys-specs} summarises the main specifications of the lensing surveys and CMB experiments considered in this section.

Figure \ref{fig:C_l_LSST_x_CMB_noise} shows various angular power spectra assuming different lensing source redshifts (colour-coded; see the figure caption) and the noise levels for different weak-lensing surveys and CMB experiments.
The latter include the $1/\sqrt{2\ell+1}$ factor related to the total number of modes at a given $\ell$ that contribute to the SNR (cf. Eq.~\eqref{eq:obs_cov}).
The left panel shows the two contributions to the total convergence field and the expected shape noise level of \textsc{lsst}, which shows that it will not be possible to detect the gravitomagnetic convergence via lensing alone \citep{Andrianomena:2014sya,Cuesta-Lazaro:2018uyv}. The middle panel shows the kSZ signal along with the lensed CMB signal, which dominates over the former down to $\ell\sim3500$, as well as the noise levels of \textsc{planck} (dashed), and of two next-generation CMB experiments; the Simons Observatory (\textsc{so}, dotted) and \textsc{cmb}-\textsc{s}4 (dot-dashed). 
We note that the kSZ effect is above the expected noise levels of the latter two CMB experiments.
Finally, the right panel shows the kSZ-$\kappa_\B$ cross spectrum and the total noise. We find that, while for \textsc{planck} the signal is almost completely dominated by the instrumental resolution on small angular scales, the situation improves considerably with the Simons Observatory and \textsc{cmb}-\textsc{s}4, in which the signal is well above the noise on scales $\ell\gtrsim3000$, which suggests that a potential future detection can be achieved on very small angular scales.
In the right panel of Fig.~\ref{fig:C_l_LSST_x_CMB_noise} we have also included the signal due to the spurious cross correlation between the ISW effect and weak lensing (colour-coded, dashed) mentioned above. This is calculated from Eq.~\eqref{eq:Cl_Theta_kappa} using the nonlinear matter power spectra at different redshifts predicted by \textsc{camb} with \textsc{halofit}. We find that, at $\ell\gtrsim3000$, this spurious signal is over one order of magnitude smaller than the gravitomagnetic signal at all redshifts, and two orders of magnitude lower at $\ell\gtrsim5000$.
Furthermore, the signal is below the noise level expected for all experiments herein considered.
Hence, in the following SNR forecast we use Eq.~\eqref{eq:obs_cov} to estimate the covariance, in which the ISW-$\kappa_\Phi$ cross-correlation is neglected.

\begin{figure*}
    \centering
    \includegraphics[width=1.0\linewidth]{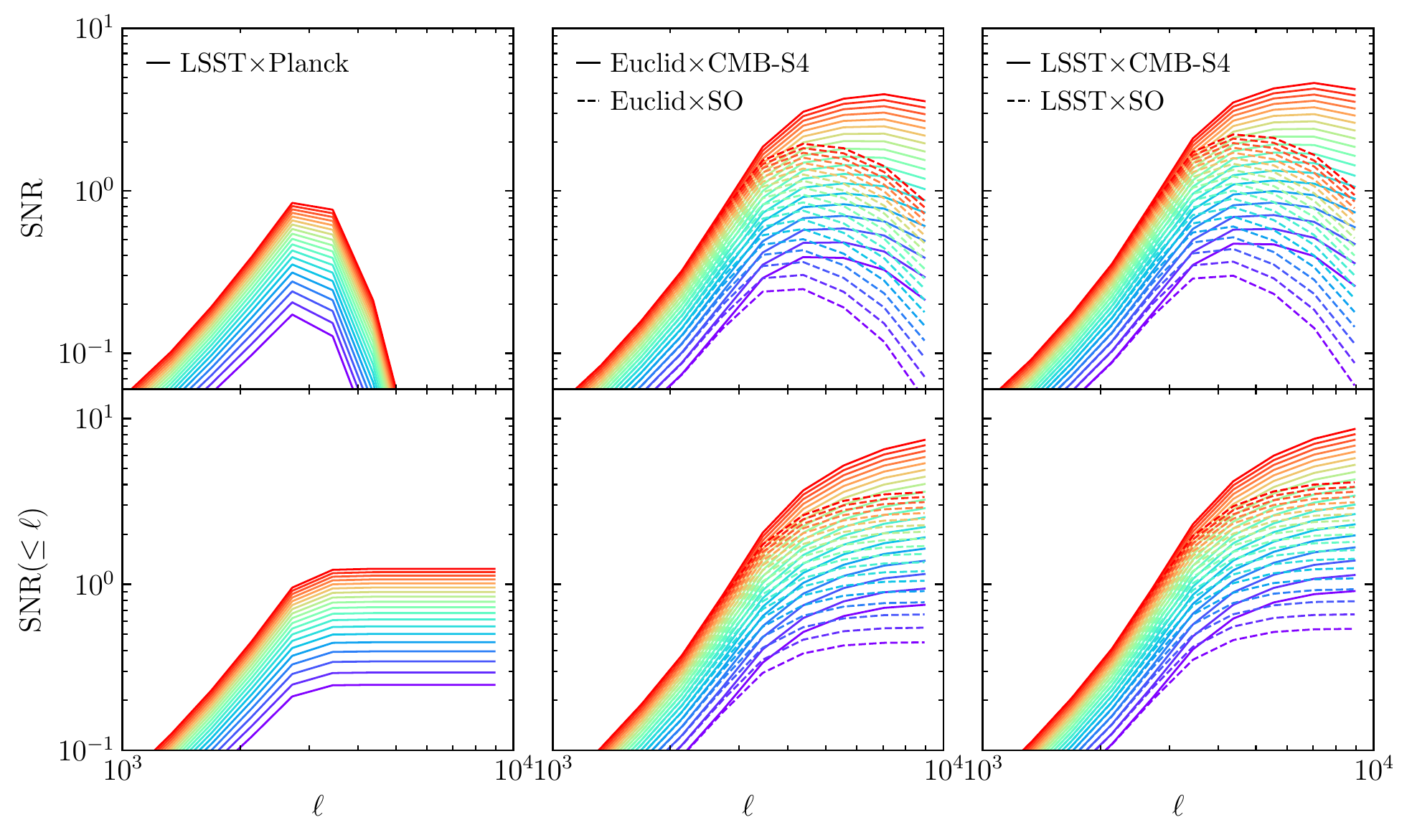}
    \caption{(Colour Online) Signal-to-noise (SNR; top panels) and cumulative SNR (bottom panels) predictions for the kSZ-$\kappa_\B$ signal via cross-correlation of different weak-lensing surveys and CMB experiments for lensing source redshifts between $z_s=0.4$ (purple, lowest amplitude) and $z_s=1.4$ (red, highest amplitude), with a separation $\Delta z_s$ equivalent to a comoving distance of $\chi=100\Mpch$.
    \textit{Left panels}: forecast for {\sc lsst} and {\sc planck}, which shows that a detection is not possible due to the angular resolution of the latter experiment.
    \textit{Middle panels}: forecast for {\sc euclid} in combination with {\sc cmb-s4} (solid) and the Simons Observatory ({\sc so}, dashed).
    \textit{Right panels}: forecast for {\sc lsst} in combination with {\sc cmb-s4} (solid) and the Simons Observatory (dashed). The angular resolution of next-generation CMB experiments may allow a significant detection of the gravitomagnetic effect.
    The results use 23 $\ell$-bins spaced logarithmically between $\ell_{\rm min}=40$ and $\ell_{\rm max}=10^4$.    
    }
    \label{fig:SNR-all-redshift}
\end{figure*}

Figure \ref{fig:SNR-all-redshift} shows the predicted SNR for different source redshifts (colour-coded; see the figure caption). In the case of cross correlating \textsc{lsst} with \textsc{planck} (left panel), we find that the instrumental resolution of \textsc{planck} is the main limiting factor, which does not allow to yield a significant detection. However, with the improved resolution of the upcoming CMB experiments such as \textsc{cmb}-\textsc{s4} and the Simons Observatory (middle and right panels), a significant detection might be achieved on small angular scales. With a lensing source redshift of $z_s=1.4$ in \textsc{lsst} (right panel), in combination with the Simons Observatory, we find that the cumulative SNR can reach around 3 (4) at $\ell\approx5000$ ($10^4$), while in the case of \textsc{cmb}-\textsc{s4} this can reach almost 5 (9) at $\ell\approx5000$ ($10^4$).
The results are similar in the case of \textsc{euclid} in combination with the two aforementioned CMB experiments (middle panel), although the SNR is slightly lower than for \textsc{lsst} due to the smaller sky coverage and mean number of galaxies expected for this survey. The results show that the majority of the cumulative SNR comes from $\ell\gtrsim2000$, and that the SNR is mainly determined by the beam size of the CMB experiment, followed by its noise level, $\Delta_T$.

From Fig.~\ref{fig:SNR-all-redshift} we can also observe that the detection SNR increases with source redshift in general (for a given CMB experiment). This is expected: as the redshift range for the LOS projection increases, the cross correlation between the gravitomagnetic lensing ($\kappa_{{\bf B}}$) and the kSZ effect ($b$) also enhances; the covariance matrix also increases, but not by as much given that $C^{\rm T}_{\ell}$ is not affected. This implies that it is possible to improve the prospect of observationally detecting the gravitomagnetic effect by using deeper lensing surveys. 
Because our high-resolution simulation does not have snapshots at even higher $z$, in this work we only have considered a limited source redshift range, and we plan to revisit this topic in the future using larger simulations.
Likewise, using a CMB lensing signal -- whose kernel peaks at $z\sim2$ -- instead of cosmic shear, may also enhance the overall lensing-CMB cross-correlation signal, and it is likely to boost the SNR. This may also have the benefit of using the lensing convergence map and the temperature map from a single CMB experiment, without a weak lensing survey. Existing data from {\sc act} \citep{Darwish2021} may provide such a possibility.

\section{Conclusion}
\label{sec:conclusion}

In this paper, we have explored the possibility of detecting the cosmological gravitomagnetic (frame-dragging) effect via cross correlation of weak-lensing convergence maps, which include the gravitomagnetic contributions, with the kSZ effect that is imprinted as a secondary anisotropy in the CMB temperature maps. The latter is chosen because -- apart from very large angular scales -- it is sourced by the rotational modes of the momentum field of matter along the LOS, just like the former effect, and at the same time is not correlated with the standard (Newtonian) component of the convergence field at the two-point level~\citep{Dore:2003ex}. Thus, the cross-correlation is able to extract the gravitomagnetic contribution from a lensing convergence map.
To model the cross-correlation signal and its covariance we have used the data from 30 Newtonian $N$-body simulations, as well as a single high-resolution, general relativistic simulation. Performing LOS projections and generating {\sc healpix} maps, we have found that small, unresolved scales close to the observer's location due to an insufficient simulation resolution can induce significant spurious variance in the cross correlation between the Newtonian component of the convergence and the kSZ effect. On the other hand, by cross correlating {\sc healpix} maps of fields taken from different realisations, such an artificial noise is not present and the covariance agrees well with the theoretical prediction of cosmic variance effects, Eq.~\eqref{eq:std-general-theory}. Then, to quantify the SNR we resort to model the signal based on the single high-resolution simulation and the Limber-approximated integral Eq. \eqref{eq:cl_xy}, and we estimate the noise by either Eq.~\eqref{eq:std-general-theory}, which includes only the effect of cosmic variance -- and allows to calculate a theoretical upper bound of the SNR -- or Eq.~\eqref{eq:obs_cov}, which also include all the major relevant effects for observations. In the former case, we find that at $z_s=0.83$, the cumulative SNR can reach $\sim15$ already at $\ell\simeq5000$, and about 30 at $\ell\simeq10^4$.

We then forecast the SNR for current CMB data from \textsc{planck}, in combination with future-weak lensing surveys such as \textsc{euclid} and \textsc{lsst}, finding that the gravitomagnetic effect cannot be robustly probed using this method as the angular resolution of \textsc{planck} does not allow to explore the small angular scales where the theoretical SNR rises most rapidly (Fig.~\ref{fig:C_l-SNR}). However, based on future CMB experiments such as the Simons Observatory and \textsc{cmb}-\textsc{s4}, our forecast shows that this effect can be detected decisively, especially with lensing sources further afield.

The result above is based on the assumption that several important late-time secondary effects on the CMB, such as the thermal SZ effect and CIB, could be reliably disentangled from the primary CMB signal, and the SNR can be degraded if such `cleaning' is not fully complete. 
We also expect that at the small scales ($\ell\gtrsim3000$) where the SNR of the effect is relatively significant, the impact of baryons on both weak lensing and CMB observables can also be significant and hence downgrade the SNR. Modelling the impact of baryonic effects on the SNR above this regime is beyond the scope of this paper and left as future work. On the other hand, given that for the kSZ effect the longitudinal modes of the momentum field are subdominant with respect to the rotational-modes contribution above $\ell\sim100$ \citep{Park:2015jea}, we do not expect them to affect our predictions. In fact SPT-SZ plus SPTpol has recently reported a $\sim 3\sigma$ measurement for kSZ, from among other dominant CMB secondary components, on the similar $\ell$-range of our interests in a 2540 deg$^2$ area \citep{SPTpol2021}. This indicates that it is promising to pick up the gravitomagnetic effect via kSZ-lensing cross-correlations.

The realistic possibility of detecting the cosmological gravitomagnetic effect with future weak-lensing surveys and CMB experiments suggests that it is worthwhile to explore the lensing-kSZ cross-correlation in the context of dark energy and modified gravity theories, in which the amplitude of both the kSZ effect can be significantly enhanced. 
For instance, in typical $f(R)$ gravity models, the magnitude of the kSZ effect can be enhanced by $\sim30\%$ at $\ell\gtrsim3000$ relative to GR~\citep[e.g.,][]{Bianchini:2015iaa,Mitchell:2020fnj}, while the magnitude of the gravitomagnetic potential can be over $40\%$ larger than in GR at $k\gtrsim2\invMpch$~\citep{Thomas:2015dfa,Reverberi:2019bov}.
Therefore, we expect to find a larger signal in these models, which could potentially be used as a new way to constrain deviations from \LCDM, for example, a non-detection can be used to place an upper bound of the strength of modified gravity. 
On the other hand, it also worthwhile to study the gravitomagnetic effects in CMB lensing and its cross correlation with CMB temperature maps. Given that the kernel of the former effect peaks at $z\sim2$, this would allow to include more signal from higher redshifts than a weak-lensing survey, and it is likely to boost the SNR. 

Another possible future work motivated by this study is the cross correlation between the kSZ effect and gravitomagnetic lensing in configuration space, where the effect is more localised. (the analysis here has been performed in Fourier space). The gravitomagnetic effect leaves the unique dipole feature in the lensing convergence field as well as in the CMB temperature map through kSZ. This may allow us to develop matched filters to single out the signal we are after. An attempt to detect the latter effects has been made recently by \citet{Tang:2020com}, where the focus is to look for a dipole feature in the lensing convergence field produced by the rotation of massive haloes. Similarly, the rotation of massive objects can produce a rotational kSZ effect detectable in future observations \citep[see, e.g.,][]{Baldi2018,Baxter:2019tze}. We leave these investigations as future work.

\section*{Acknowledgements}

CB-H is supported by the Chilean National Agency of Research and Development (ANID) through grant CONICYT/Becas-Chile (No.~72180214). BL is supported by the European Research Council (ERC) through ERC starting Grant No.~716532, and STFC Consolidated Grant (Nos.~ST/I00162X/1, ST/P000541/1). YC acknowledges the support of the Royal Society through a University Research Fellowship and an Enhancement Award. We thank Tianyi Yang for useful comments. We also thank the anonymous referee for their valuable comments and suggestions, which helped to improve the original manuscript.

This work used the DiRAC@Durham facility managed by the Institute for Computational Cosmology on behalf of the STFC DiRAC HPC Facility (\url{www.dirac.ac.uk}). The equipment was funded by BEIS via STFC capital grants ST/K00042X/1, ST/P002293/1, ST/R002371/1 and ST/S002502/1, Durham University and STFC operation grant ST/R000832/1. DiRAC is part of the UK National e-Infrastructure.

\section*{Data Availability}

For access to the simulation data please contact CB-H.



\bibliographystyle{mnras}
\bibliography{main} 



\appendix

\section{The kinetic Sunyaev-Zel'dovich (kSZ) effect}
\label{sect:kSZ}

CMB photons can interact with fast-moving free electrons in the intergalactic medium (IGM) via inverse Compton scattering, which subsequently changes their energy and imprints a secondary CMB anisotropy known as the kinetic Sunyaev-Zel'dovich (kSZ) effect \citep{kSZ:1980}. The temperature fluctuation along the line-of-sight (LOS) vector $\hat{\bf n}$ due to this effect can be described by the following LOS integral,
\begin{equation}\label{eq:b_ksz1}
    b(\hat{\bf n}) \equiv -\frac{\Delta T(\hat{\bf n})}{\bar{T}} = \int{\rm d}\tau e^{-\tau}\frac{\hat{\mathbf{n}}\cdot\mathbf{v}}{c} = \sigma_{\rm T}\int{\rm d}le^{-\tau}\frac{n_{\rm e}v_{\rm r}}{c}\,,
\end{equation}
in which $\bar{T}$ is the mean temperature of the CMB, $\sigma_{\rm T}$ and $\tau$ are respectively the Thomson scattering cross section and optical depth, $c$ is the speed of light, $n_{\rm e}$ is the number density of free electrons, and $v_r=\mathbf{v}\cdot\hat{\mathbf{n}}$ is the LOS component of the electron velocity field. 

Since Eq.~\eqref{eq:b_ksz1} is an effect integrated from $z=0$ to the last scattering surface, $z\approx 1100$, the kSZ signal has two distinct contributions, one coming from the post-reinoisation epoch, in which the IGM is nearly fully ionised and the electron density field closely follows the density field of baryons; and the contribution from the epoch of reionisation, where $n_{\rm e}$ suffers strong temporal and spatial variations. As the goal of this paper is to study the cross correlation of the kSZ effect with a weak lensing survey such as {\sc lsst} and {\sc euclid}, throughout the present analysis we restrict our attention to the post-reionisation kSZ signal.

The specific ionised momentum field of the ionised medium can be defined as
\begin{equation}
    \mathbf{q} \equiv \chi_{\rm e}(1+\delta)\mathbf{v}\,,
\end{equation}
where $\delta$ is the baryon density contrast, and
\begin{equation}\label{eq:ionised_fraction}
    \chi_{\rm e} \equiv \frac{n_{\rm e}}{n_{\rm H}+2n_{\rm He}}\,,
\end{equation}
denotes the ionised fraction, with $n_{\rm H}$, $n_{\rm He}$ being the number densities for hydrogen and helium, respectively. Also, defining 
\begin{equation}
    \bar{n}_{{\rm e},0} \equiv n_{{\rm H},0} + n_{{\rm He},0}\,,
\end{equation}
Eq.~\eqref{eq:b_ksz1} can be rewritten as
\begin{equation}\label{eq:b_ksz}
    b = \frac{\sigma_{\rm T}\bar{n}_{{\rm e},0}}{c}\int{\rm d}\chi\frac{1}{a^2} e^{-\tau}\mathbf{q}\cdot\hat{\mathbf{n}}\,,
\end{equation}
where $\chi$ is the comoving distance along the LOS. Using the Fourier transform of Eq.~\eqref{eq:b_ksz}, we get
\begin{equation}\label{eq:b_fourier1}
    b = \frac{\sigma_{\rm T}\bar{n}_{{\rm e},0}}{c}\int{\rm d}\chi\frac{1}{a^2(\chi)} e^{-\tau}\int\frac{{\rm d}^3\mathbf{k}}{(2\pi)^3}\left[\hat{\mathbf{n}}\cdot\tilde{\mathbf{q}}(\mathbf{k},\chi)\right]e^{-i\chi\mathbf{k}\cdot{\hat{\mathbf{n}}}}\,, 
\end{equation}
where ${\bf k}$ is the wavevector, $i$ is the imaginary number unit, and $\tilde{\bf q}$ is the momentum vector in Fourier space. One can decompose $\tilde{\bf q}$ into a longitudinal (scalar) and a rotational (vector) part:
\begin{equation}
    \tilde{\mathbf{q}} = \tilde{\bf q}_\parallel + \tilde{\bf q}_\perp, \quad {\rm with}~~\tilde{\bf q}_\parallel = (\tilde{\bf q}\cdot\hat{\bf k})\hat{\bf k}\,,
\end{equation}
where $\hat{\bf k}$ is the unit vector in the direction of the wavevector. Substituting this into Eq.~\eqref{eq:b_fourier1} gives \citep[][]{Park:2013mv} 
\begin{align}\label{eq:b_fourier2}
    b =& \frac{\sigma_{\rm T}\bar{n}_{{\rm e},0}}{c}\int{\rm d}\chi\frac{1}{a^2(\chi)} e^{-\tau}\times \\
    &\int\frac{{\rm d}^3\mathbf{k}}{(2\pi)^3}\left[x\tilde{q}_\parallel({\bf k},\chi) - \cos\left(\phi_{\hat{\bf q}}-\phi_{\hat{\bf n}}\right)\sqrt{1-x^2}\tilde{q}_\perp({\bf k},\chi)\right]e^{-ik\chi x}\,, \nonumber
\end{align}
where $\tilde{q}_\parallel=|\tilde{\bf q}_\parallel|$, $x\equiv\hat{\bf k}\cdot\hat{\bf n}$, $\phi_{\hat{\bf q}}$ and $\phi_{\hat{\bf n}}$ are respectively the angle between ${\bf q}, \hat{\bf n}$ and ${\bf k}$. The exponential function in the integral represents a fast oscillation along the LOS, which means that the integrand cancels out, leading to negligible integral result. There are two exceptions to this: (1) if $k\rightarrow0$, or (2) if $x\rightarrow0$. (1) represents a long-wave mode which has a small amplitude and therefore contributes little to the integral anyway. (2) represents a case where ${\bf k}$ is perpendicular to the LOS, $\hat{\bf n}$. In other words, only the ${\bf k}$ modes that are perpendicular to the LOS contribute to the kSZ effect non-negligibly. But in this case we can see from Eq.~\eqref{eq:b_fourier2} that the first term in the brackets vanishes since $x\rightarrow0$, and therefore only the rotational momentum field $\tilde{q}_\perp$ remains, giving \citep[e.g.,][]{Park:2013mv}
\begin{align}\label{eq:b_fourier}
    b \simeq& -\frac{\sigma_{\rm T}\bar{n}_{{\rm e},0}}{c}\int{\rm d}\chi\frac{1}{a^2(\chi)} e^{-\tau}\times \\
    &\int\frac{{\rm d}^3\mathbf{k}}{(2\pi)^3}\cos\left(\phi_{\hat{\bf q}}-\phi_{\hat{\bf l}}\right)\sqrt{1-x^2}\tilde{q}_\perp({\bf k},\chi)e^{-ik\chi x}\,. \nonumber
\end{align}
With some lengthy derivation \cite[see, e.g., Appendix A of][]{Park:2013mv}, one gets the following expression for the kSZ ($b$) angular power spectrum
\begin{equation}
    C^{b}_\ell = \frac{1}{2}\left[\frac{\sigma_{\rm T}\bar{n}_{{\rm e},0}}{c}\right]^2\int{\rm d}\chi\frac{1}{\chi^2a^4(\chi)}e^{-2\tau}P_{q_\perp}\left(k=\frac{\ell}{\chi},\chi\right)\,,
\end{equation}
where $P_{q_\perp}$ is the 3D power spectrum of $q_\perp$, the rotational momentum field. Assuming that the velocity field is completely longitudinal, as it is the case for a pressureless perfect fluid, $P_{q_\perp}$ appears only at second order through the following convolution:
\begin{align}\label{eq:P_q_perp1}
    P_{q_\perp}(k,z)& = \int\frac{{\rm d}^3{\bf k}'}{(2\pi)^3}(1-\mu^2) \nonumber \\
    &\left[P_{\delta\delta}\left(|{\bf k}-{\bf k}'|\right)P_{vv}(k')-\frac{k'}{|{\bf k}-{\bf k}'|}P_{\delta{v}}\left(|{\bf k}-{\bf k}'|\right)P_{\delta{v}}(k')\right]\,,
\end{align}
where $\mu=\hat{\bf k}\cdot\hat{\bf k}'$. If we define
\begin{equation}
    w \equiv k'/k, \quad u \equiv |{\bf k}-{\bf k}'|/k\,,
\end{equation}
then 
\begin{equation}
    \mu = \frac{1+w^2-u^2}{2w}\,,
\end{equation}
and Eq.~\eqref{eq:P_q_perp1} can be recast as
\begin{align}\label{eq:ksz_spectrum}
    P_{q_\perp}(k,z) & = \frac{k^3}{4\pi^2}\int^\infty_0{\rm d}w \nonumber \\
    &\int^{1+w}_{|1-w|}{\rm d}u \Pi'\left[P_{\delta\delta}\left(ku\right)P_{vv}(kw)-\frac{w}{u}P_{\delta{v}}\left(ku\right)P_{\delta{v}}(kw)\right]\,,
\end{align}
where
\begin{equation}
    \Pi' \equiv uw\frac{4w^2-(1+w^2-u^2)}{4w^2}\,.
\end{equation}

Given that  the gravitomagnetic field is sourced by the rotational modes of the momentum field through Eq. \eqref{eq:momentum-constraint}, Eq. \eqref{eq:ksz_spectrum} has a similar form to the gravitomagnetic potential power spectrum, which in the case of a pressureless perfect fluid is given by, e.g., \citet{Lu:2007cj}
\begin{align}\label{eq:B_spectrum1}
    \Delta_{\bf B}(k)& = \frac{9\Omega_m^2H_0^2}{2a^2c^2k^2}\int^\infty_0{\rm d}w \nonumber \\
    &\int^{1+w}_{|1-w|}{\rm d}u\Pi\left[\Delta_{\delta\delta}(ku)\Delta_{vv}(kw) - \frac{w}{u}\Delta_{\delta{v}}(ku)\Delta_{\delta{v}}(kw)\right]\,,
\end{align}
where the dimensionless power spectrum $\Delta_{ab}$ is defined as
\begin{equation}\label{eq:Delta}
    \Delta_{ab} \equiv \frac{k^3}{2\pi^2}P_{ab}\,,
\end{equation} 
with $a,b$ standing for two fields, and
\begin{equation}
    \Pi \equiv \frac{1}{u^2w^2}\frac{4w^2-(1+w^2-u^2)}{4w^2}\,.
\end{equation}
Indeed, substituting Eq.~\eqref{eq:Delta} into Eq.~\eqref{eq:B_spectrum1} gives
\begin{align}\label{eq:B_spectrum}
    P_{\bf B}(k)& = \frac{k}{2\pi^2}\frac{9\Omega_m^2H_0^2}{2a^2c^2}\int^\infty_0{\rm d}w \nonumber \\
    &\int^{1+w}_{|1-w|}{\rm d}u\Pi'\left[P_{\delta\delta}(ku)P_{vv}(kw) - \frac{w}{u}P_{\delta{v}}(ku)P_{\delta{v}}(kw)\right]\,.
\end{align}
which differs from \eqref{eq:ksz_spectrum} only in the prefactor (and the fact that here the density and velocity power spectra are for all matter, rather than free electrons only; the two are closely related) including a $k^2$. 

Following the Appendix A of \citet{Park:2013mv}, we can derive a Limber-integral expression (which has been used extensively in this paper) for the cross angular power spectrum between $\kappa_{\bf B}$ and the kSZ effect ($b$). Given the above mathematical similarity between the two effects, the detailed steps will not be repeated here to be concise as the calculation is similar to the derivation of the kSZ auto-power spectrum. For generality, consider two 2D fields $X(\hat{\bf n})$ and $Y(\hat{\bf n})$ which are both related to the projection of the LOS momentum field ${\bf q}\cdot\hat{\bf n}$:
\begin{equation}\label{eq:X_fourier1}
    X,Y(\hat{\bf n}) = \int{\rm d}z{K}_{X,Y}(\chi)\int\frac{{\rm d}^3\mathbf{k}}{(2\pi)^3}\left[\hat{\mathbf{n}}\cdot\tilde{\mathbf{q}}\left(\mathbf{k},z(\chi)\right)\right]e^{-i\chi\mathbf{k}\cdot{\hat{\mathbf{n}}}}\,, 
\end{equation}
where $K_{X,Y}(s)$ are respectively the LOS projection kernels for observables $X$ and $Y$, which are functions of the comoving distance $\chi$. The cross angular power spectrum between $X$ and $Y$, $C^{XY}_\ell$, is defined as
\begin{equation}\label{eq:Cl_def}
    C_\ell^{XY}\delta_{\ell\ell'}\delta_{mm'} = \left\langle{a}^{X}_{\ell m}a^{Y\ast}_{\ell'm'}\right\rangle\,,
\end{equation}
where $\langle\cdots\rangle$ denotes ensemble average, $\ast$ denotes the complex conjugate, and $a_{\ell{m}}^{X,Y}$ are the spherical harmonic decomposition coefficients for $X$ and $Y$,
\begin{equation}\label{eq:sph_harmonic_coeffs}
a^{X,Y}_{\ell m} = \int{\rm d}^2\hat{\bf n}X,Y(\hat{\bf n})Y_\ell^{m\ast}(\hat{\bf n})\,,
\end{equation}
with $Y^m_\ell$ being the spherical harmonic function of degree $\ell$ and order $m$. Hence, $C^{XY}_{\ell}$ can be expressed as a weighted LOS integration of the 3D power spectrum of the rotational component of the LOS momentum field ${\bf q}\cdot\hat{\bf n}$, $P_{q_\perp}$, as:
\begin{equation}
    C^{XY}_{\ell} \simeq \frac{1}{2}\int{\rm d}\chi\chi^{-2}K_X(\chi)K_{Y}(\chi)P_{q_\perp}\left(k=\frac{\ell}{\chi},z(\chi)\right)\,.
\end{equation}

\section{The ISW-weak lensing cross correlation}
\label{sect:isw-wl-x}

In this appendix we derive an expression for the cross angular power spectrum between the integrated Sachs-Wolfe effect and weak lensing convergence. For simplicity, we assume again a single lensing source redshift $z_s$. The derivation follows the appendix of \citet{Cai:2008sm}, see also \citep{Seljak1996, Smith2009, Nishizawa2014}.
The CMB temperature fluctuation induced by the ISW effect along the LOS direction $\hat{\mathbf{n}}$ is given by
\begin{equation}
    \Theta \equiv \frac{\Delta T\left(\hat{\mathbf{n}}\right)}{\bar{T}} = \frac{2}{c^2}\int^{t_0}_{t_{\rm LSS}}\dot{\Phi}\left(t,\chi(t)\hat{\mathbf{n}}\right)\mathrm{d}t\,,
\end{equation}
where $\bar{T}$ is the mean CMB temperature, $t$ is the cosmic time, $\chi$ is the comoving distance along the LOS, and $\dot{\Phi}$ the time derivative of the gravitational potential $\Phi$; $t_0$ and $t_{\rm LSS}$ are respectively the values of $t$ at the observer (today) and the last-scattering surface. The spherical harmonic coefficients of $\Theta$, defined in the same way as in Eq.~\eqref{eq:sph_harmonic_coeffs}, can be expressed as
\begin{equation}\label{eq:a_lm_Theta}
    a^{\Theta}_{\ell{m}} = \frac{i^\ell}{\pi^2c^2}{\int^{t_0}_{t_{\rm LSS}}{\rm d}t}\int\mathrm{d}^3\mathbf{k}\dot{\Phi}(\mathbf{k},a(\chi))j_\ell(k\chi)Y_{\ell}^{m\ast}\left(\hat{\mathbf{k}}\right)\,,
\end{equation}
where $i$ is the imaginary number unit, the scale factor $a$ is written as a function of the comoving LOS distance $\chi$, $\chi_{s}\equiv{\chi\left(z_s\right)}$ the comoving distance of the source, and $\Phi(\mathbf{k})$ the Fourier transform of $\Phi({\boldsymbol{\chi}})$:
\begin{equation}
    \Phi\left(\mathbf{k}\right) = \frac{1}{(2\pi)^3}\int\mathrm{d}^3\boldsymbol{\chi}\Phi(\boldsymbol{\chi})\exp(i\mathbf{k}\cdot\boldsymbol{\chi})\,.
\end{equation}
In deriving Eq.~\eqref{eq:a_lm_Theta}, we have used the spherical harmonic expansion of a plane wave:
\begin{equation}
    \exp\left(i\mathbf{k}\cdot\boldsymbol{\chi}\right) = 4\pi\sum_{\ell{m}}i^\ell j_{\ell}(k\chi)Y_{\ell}^{m\ast}(\hat{\mathbf{k}})Y_{\ell}^m(\hat{\mathbf{n}})\,,
\end{equation}
and the orthonormality of the spherical harmonics:
\begin{equation}
    \int_{\Omega}\mathrm{d}^2\hat{\mathbf{n}}Y_{\ell}^{m\ast}(\hat{\mathbf{n}})Y_{\ell'}^{m'}(\hat{\mathbf{n}}) = \delta_{\ell'\ell}\delta_{m'm}\,,
\end{equation}
where $\Omega$ denotes the solid angle, and $\delta_{\ell'\ell}$ and $\delta_{m'm}$ are the Kronecker deltas. 
The spherical harmonic expansion coefficient of the weak lensing convergence field, $\kappa$, 
\begin{equation}
    \kappa\left(\hat{\mathbf{n}}\right) = \frac{3H_0^2\Omega_m}{2c^2}\int_0^{\chi_{s}}\mathrm{d}\chi\frac{(\chi_{s}-\chi)\chi}{\chi_{s}}\frac{\delta}{a}\,,
\end{equation}
can be similarly obtained, as 
\begin{equation}
    a^{\kappa}_{\ell{m}} = \frac{3H_0^2\Omega_m}{4\pi^2c^2}i^\ell\int^{\chi_{s}}_0\mathrm{d}\chi\frac{(\chi_{s}-\chi)\chi}{\chi_{s}a(\chi)}\int\mathrm{d}^3\mathbf{k}\delta(\mathbf{k},a(\chi))j_\ell(k\chi)Y_{\ell}^{m\ast}\left(\hat{\mathbf{k}}\right)\,,
\end{equation}
where $\delta(\mathbf{k},a)$ is the Fourier transform of the density contrast field at scale factor $a$. 
Using the definition of the cross angular power spectrum given in Eq.~\eqref{eq:Cl_def}, we get, after some lengthy but straightforward derivation,
\begin{equation}\label{eq:Cl_Theta_kappa_1}
    C^{\Theta\kappa}_{\ell} = \frac{3H_0^2\Omega_m}{c^5}\int_0^{\chi_{s}}{\rm d}\chi\frac{\chi_{s}-\chi}{\chi_{s}\chi}P_{\dot{\Phi}\delta}\left(k=\frac{\ell}{\chi},a(\chi)\right)\,,
\end{equation}
where we have used $c{\rm d}t=-a(t){\rm d}\chi$, and the 3D cross power spectrum between $\dot{\Phi}$ and $\delta$, $P_{\dot{\Phi}\delta}$, is given by
\begin{equation}
    (2\pi)^3\delta^{(3)}\left(\mathbf{k}-\mathbf{k}'\right)P_{\dot{\Phi}\delta}(k,a) = \left\langle\dot{\Phi}^\ast(\mathbf{k},a)\delta(\mathbf{k}',a)\right\rangle\,,
\end{equation}
where $\delta^{(3)}(\mathbf{k}-\mathbf{k}')$ is the 3D Dirac $\delta$ function. 
To evaluate $P_{\dot{\Phi}\delta}$, we make use of the Fourier-space Poisson equation, 
\begin{equation}
    -k^2\Phi(\mathbf{k}) = \frac{3}{2}H_0^2\Omega_m\frac{\delta(\mathbf{k})}{a}\,,
\end{equation}
to get the derivative of $\Phi(\mathbf{k})$, as
\begin{equation}
    \dot{\Phi}(\mathbf{k}) = -\frac{3}{2}\left(\frac{H_0}{k}\right)^2\Omega_m\left[\frac{\dot{\delta}(\mathbf{k})}{a}-\frac{H}{a}\delta(\mathbf{k})\right]\,,
\end{equation}
with $H=\dot{a}/a$ being the Hubble expansion rate at $a$. Therefore, we have
\begin{equation}
    \left\langle\dot{\Phi}^\ast(\mathbf{k})\delta(\mathbf{k}')\right\rangle = -\frac{3}{2}\left(\frac{H_0}{k}\right)^2\Omega_m\left\langle\frac{\dot{\delta}(\mathbf{k})}{a}-\frac{H}{a}\delta(\mathbf{k}),\delta(\mathbf{k}')\right\rangle\,,
\end{equation}
and 
\begin{equation}
    P_{\dot{\Phi}\delta} = -\frac{3a}{4}\left(\frac{H_0}{k}\right)^2\Omega_m\frac{\mathrm{d}}{\mathrm{d}t}\left(a^{-2}P_{\delta\delta}(k,a)\right)\,,
\end{equation}
where $P_{\delta\delta}$ is the matter power spectrum.
Using the above relation, Eq.~\eqref{eq:Cl_Theta_kappa_1} can be simplified as
\begin{equation}\label{eq:Cl_Theta_kappa}
     C^{\Theta\kappa}_\ell= \frac{9H_0^4\Omega_m^2}{4c^4\ell^2}\int^{z_s}_0\mathrm{d}z\frac{\left(\chi_{s}-\chi\right)\chi}{\chi_{s}}\frac{\mathrm{d}}{\mathrm{d}z}\left[(1+z)^{2}P_{\delta\delta}\left(k=\frac{\ell}{\chi},z\right)\right]\,.
\end{equation}
where we have changed the integration variable and time derivatives to $z$.
The $C^{\Theta\kappa}_\ell$ cross angular power spectrum is shown in the right panel of Fig.~\ref{fig:C_l_LSST_x_CMB_noise}, for which Eq.~\eqref{eq:Cl_Theta_kappa} is evaluated using the nonlinear matter power spectra at different redshifts predicted by \textsc{camb} with \textsc{halofit}. {We find qualitatively similar result to the cross spectrum between the ISW effect and galaxies \citep[e.g., Fig.~5 of][]{Cai:2008sm}.}

\bsp	
\label{lastpage}
\end{document}